\documentclass[letterpaper,12pt]{article}
\usepackage{mathptmx}
\usepackage{amsmath}
\usepackage{amsfonts}
\usepackage{amsthm}
\usepackage{mathtools}
\usepackage{amssymb}
\usepackage[pdftex]{graphicx}
\usepackage{stmaryrd}
\usepackage{float}
\usepackage{mathrsfs}
\usepackage{bbold}
\usepackage{caption}
\usepackage{BOONDOX-cal}
\usepackage{hyperref}

\usepackage{graphics}

\usepackage[authoryear,comma,sectionbib, sort]{natbib} 

\usepackage{subfig}

\DeclareMathAlphabet{\mathcall}{U}{BOONDOX-cal}{m}{n}
\SetMathAlphabet{\mathcall}{bold}{U}{BOONDOX-cal}{b}{n}

\usepackage{tikz}
\usetikzlibrary{arrows,shapes}
\tikzstyle{var}=[draw=none,fill=none]
\tikzstyle{unobsvar}=[circle, draw,fill=none]
\tikzstyle{edge} = [draw,thick,->]
\tikzstyle{edge2} = [draw,thick, dashed,->]  
\usetikzlibrary{arrows,shapes}
\tikzstyle{edge2gris} = [draw,thick, dashed,->,CadetBlue]
\tikzstyle{edgebleu} = [draw,thick,->, NavyBlue]
\tikzstyle{edgerouge} = [draw,thick,->, OrangeRed]
\tikzstyle{edgeBleu} = [draw,thick,dotted,->, NavyBlue]
\tikzstyle{edgeRouge} = [draw,thick, dotted, ->, OrangeRed]
\tikzstyle{edgegris} = [draw,thick, ->, gray]

\renewcommand{\P}{{\rm I}\kern-0.14em{\rm P}}
\newcommand{\p}{{\rm I}\kern-0.18em{\rm P}}
\newcommand{\E}{{\rm I}\kern-0.14em{\rm E}}

\newtheorem{Theorem}{Theorem}

\def\1{{\rm 1\mskip-4.4mu l}}

\renewcommand{\P}{\mathbb{P}}

\usepackage{color}

\newcommand{\magenta}[1]{\textcolor{magenta}{#1}}

\makeatletter
\newcommand*{\indep}{%
  \mathbin{%
    \mathpalette{\@indep}{}%
  }%
}
\newcommand*{\nindep}{%
  \mathbin{%                   % The final symbol is a binary math operator
    \mathpalette{\@indep}{\not}% \mathpalette helps for the adaptation
  }%
}
\newcommand*{\@indep}[2]{%
  \sbox0{$#1\perp\m@th$}%        box 0 contains \perp symbol
  \sbox2{$#1=$}%                 box 2 for the height of =
  \sbox4{$#1\vcenter{}$}%        box 4 for the height of the math axis
  \rlap{\copy0}%                 first \perp
  \dimen@=\dimexpr\ht2-\ht4-.2pt\relax
  \kern\dimen@
  {#2}%
  \kern\dimen@
  \copy0 %                       second \perp
} 

\usepackage{ulem}

\author{Lola Étiévant*, Vivian Viallon}
\title{Causal inference under over-simplified longitudinal causal models}
\date{}

\pdfoutput=1

\begin{document}

\maketitle

\noindent *Corresponding author: Lola Étiévant, Institut Camille Jordan, Villeurbanne 69622, France, E-mail: lola.etievant@gmail.com. https://orcid.org/0000-0001-7562-3550\\

\noindent Vivian Viallon, Nutritional Methodology and Biostatistics, International Agency for Research on Cancer, Lyon 69372, France, E-mail: viallonv@iarc.fr\\

\noindent The final published article and its Supplementary Materials are available at the \textcolor{red}{\href{https://www.degruyter.com/document/doi/10.1515/ijb-2020-0081/pdf}{International Journal of Biostatistics}} website. 

\section*{Abstract}

Many causal models of interest in epidemiology involve longitudinal exposures, confounders and mediators. However, repeated measurements are not always available or used in practice, leading analysts to overlook the time-varying nature of exposures and work under over-simplified causal models. 
Our objective is to assess whether - and how - causal effects identified under such misspecified causal models relates to true causal effects of interest.
We derive sufficient conditions ensuring that the quantities estimated in practice under over-simplified causal models can be expressed as weighted averages of longitudinal causal effects of interest.
Unsurprisingly, these sufficient conditions are very restrictive, and our results state that the quantities estimated in practice should be interpreted with caution in general, as they usually  do  not  relate to any longitudinal causal effect of interest. Our simulations further illustrate that the bias between the quantities estimated in practice and the weighted averages of longitudinal causal effects of interest can be substantial.  Overall, our results confirm the need for repeated measurements to conduct proper analyses and/or the development of sensitivity analyses when they are not available. \\

\textbf{Keywords:} Causal inference, longitudinal model, identifiability, structural causal model.

\section{Introduction}

Etiologic epidemiology is concerned with the study of potential causes of chronic diseases based on observational data. Over the years, it has been successful in the identification of links between several lifestyle exposures and the risk of cancer for example. Remarkable examples are tobacco smoke, alcohol and obesity that are now established risk factors for the development of a number of site-specific cancers \citep{AGU_BON, BAG_ROT, LAUBY}. Moreover, an accumulating body of biomarker measurements and -omics data provide important opportunities for investigating biological mechanisms potentially involved in cancer development. For example, cancer epidemiology is increasingly concerned by the study of the carcinogenic role of inflammation, insulin resistance and sex steroids hormones \citep{bradbury2019circulating, chan2011inflammatory, dossus2013hormonal}. 

As they are based on observational data, the causal validity of these analyses relies on strong assumptions, which have been formally described in the causal inference literature \citep{H_R, pearl2009statsurvey, pearl_book, rosenbaum1983, robins1986}. The very first assumption underlying most causal analyses is that the causal model is correctly specified. Most often, {\it e.g.}, when studying lifestyle exposures such as tobacco smoke, alcohol and obesity, but also biomarkers, the true causal model involves time-varying risk factors. Valid causal inference under such longitudinal causal models usually requires repeated measurements for the time-varying variables \citep{G_Formula, VDW_Book, VDW_TCH}. However, repeated measurements are still rarely available in large prospective epidemiological studies; exceptions include electronic health records, but their analyses raise other challenges (see Section \ref{sec:Discu}). 
Issues arising when ignoring the time-dependant nature of the exposures have been described in the literature \citep{AALEN, article2, article1}. Moreover, general results on the identifiability of causal effects in the presence of unobserved variables can be used to study the identifiability of causal effects of interest when ignoring the time-varying nature of exposures \citep{ShpitserPearl2006a, TianPearl2002, HuangValtorta2006, TianPearl2003}. 

However, little is known about the possible relationship between estimates derived under oversimplified longitudinal causal models and causal quantities of interest under the true longitudinal causal model. Filling this gap is the main objective of the present work. 
Specifically, we derive sufficient conditions that guarantee that the quantity estimated in practice when working under over-simplified causal models is related to longitudinal causal effects of interest. Unsurprisingly, these sufficient conditions are very restrictive, and are met under very simple causal models only. Through numerical examples, we show that the magnitude of the bias between the quantities estimated in practice and longitudinal causal effects of interest can be substantial. Overall, our results raise the need for repeated measurements to conduct proper analyses and/or the development of sensitivity analyses when they are not available.

The rest of the article is organized as follows. The notation and framework considered in this work is introduced in Section \ref{Sec:Notation_Setting}. In section \ref{Sec:MainRes}, we present our main result and discuss its practical implications. Section \ref{sec:Illustr_SV} is devoted to numerical illustrations. Concluding remarks are provided in Section \ref{sec:Discu}. Technical derivations are presented in the Appendix.

\section{Notation}\label{Sec:Notation_Setting}

\subsection{The true longitudinal causal model}\label{Subsec:Notation_TrueModel}

We consider standard longitudinal causal models \citep{G_Formula, VDW_Book}, where time-varying exposures, including the exposure of interest, but also possibly mediators and confounders, are observable at discrete times over the time-window $\llbracket 1 ; T \rrbracket := \{1, \ldots, T\}$, for some time $T > 1$. For any $t\in\llbracket 1 ; T \rrbracket$, we denote by $X_t$ the exposure of interest at time $t$, by $\bar X_t = (X_1, X_2, \dots , X_t)$ the exposure profile until time $t$, while $\bar x_t$ stands for a specific (fixed) profile for the exposure of interest \citep{VDW_Book}. In addition, $\underline X_{t_1}^{t_2} = (X_{t_1}, X_{t_1+1}, \dots , X_{t_2})$ denotes the exposure profile from time $t_1$ to time $t_2$, $ 1 \leq t_1 \leq t_2 \leq T$. We use similar notation for auxiliary factors $(Z_t)_{t}$, which may include ``pure mediator'' processes $(M_t)_{t}$ and confounder processes $(W_t)_{t}$ possibly affected by the exposure of interest (as in Figure \ref{fig:example_longitudinal_XMW}). By ``pure mediator'' process, we mean that $M_t$ does not affect $X_{t_1}$, for any $t, t_1\in\llbracket 1 ; T \rrbracket$, so that variables $(M_t)_t$ only act as mediators (and not as confounders) in the relationship between the exposure of interest and the outcome. Similarly, a ``pure confounder'' refers to a variable, or process, that possibly affects, but is not affected by, the exposure. We denote by $Y$ the outcome of interest measured at time $T$. For example $(X_t)_{t}$ could represent  body mass index (BMI) at different ages, $Y$ could represent cancer development by a given age, while the auxiliary multivariate variable $(Z_t)_{t}$ could represent alcohol intake, physical activity and dietary exposures at different ages. For illustration, we will mostly consider the example of the longitudinal causal model $(L.1)$ provided in Figure \ref{fig:example_longitudinal_XMW}, as well as special cases of this model. Unless otherwise stated, we further assume that all variables are binary to simplify our notation. 

\begin{center}

\begin{figure}
\begin{minipage}[c]{1\linewidth}
\begin{center}
\begin{tikzpicture}[scale=0.8, auto,swap]
\node[var] (W2)at(2,0){$W_1$};
\node[var] (W3)at(4,0){$W_2$};
\node[var] (Wdot)at(5.5,0){$\dots$};
\node[var] (Wt)at(7,0){$W_{T}$};
\node[var] (X2)at(2,-1.5){$X_1$};
\node[var] (X3)at(4,-1.5){$X_2$};
\node[var] (Xdot)at(5.5,-1.5){$\dots$};
\node[var] (Xt)at(7,-1.5){$X_{T}$};
\node[var] (Y)at(8.75,-0.75){$Y$};

\draw[edge] (X2)--(X3);
\draw[edge] (X2)--(W3);
\draw[edge] (X2).. controls (3.5, -0.85)..(Wt);
\draw[edge] (X3)--(Wt);

\draw[edge] (W2)--(W3);
\draw[edge] (W2)--(X2);
\draw[edge] (W2)--(X3);
\draw[edge] (W2).. controls (3.5, -0.75)..(Xt);
\draw[edge] (W3)--(Xt);

\draw[edge] (W3)--(X3);
\draw[edge] (W3)--(Wdot);
\draw[edge] (Wdot)--(Wt);
\draw[edge] (Wt)--(Xt);
\draw[edge] (Wt)--(Y);

\draw[edge] (X3)--(Xdot);
\draw[edge] (Xdot)--(Xt);
\draw[edge] (Xt)--(Y);

\draw[edge] (W2).. controls (5.1,0.75) ..(Wt);
\draw[edge] (W2).. controls (6,1.15) ..(Y);

\draw[edge] (W3).. controls (5.1,0.5) ..(Wt);
\draw[edge] (W3).. controls (7.2,1.15) ..(Y);

\draw[edge] (X2).. controls (5.1,-2.2) ..(Xt);
\draw[edge] (X2).. controls (6,-2.5) ..(Y);
%\draw[edge] (X2).. controls (5,-3.5) and (8.5,-3.5)..(Y);

\draw[edge] (X3).. controls (5.1,-1.9) ..(Xt);
\draw[edge] (X3).. controls (7.2,-2.5) ..(Y);
\end{tikzpicture}
\end{center}
\end{minipage}\hfill

\vspace{0.1cm}

\begin{minipage}[c]{1\linewidth}
\centering
($L.1$)
\end{minipage}

    \caption{(\textit{L.1}) Example of longitudinal causal model with a time-varying confounder $(W_t)_{t\geq 1}$ possibly affected by the time-varying exposure of interest $(X_t)_{t\geq 1}$.  
    }\label{fig:example_longitudinal_XMW}
\end{figure}
\end{center}

For any pair of variables $(V ,U)$ and any potential value $u$ of $U$, we denote by $V^{U = u}$ the counterfactual variable corresponding to variable $V$ that would have been observed in the counterfactual world following the hypothetical intervention $do(U = u)$.  We work under the setting of Structural Causal Models \citep{pearl2009statsurvey}, which especially entails that consistency conditions hold: for instance, $U=u$ implies $V = V^{U = u}$. In addition, some positivity conditions \citep{rosenbaum1983}, specified in the conditions of Theorem \ref{Theo_SV_Weak} in Section \ref{Sec:MainRes}, will be assumed to hold.  For any possibly counterfactual random variables $V$ and $U$, and any causal model $(Mod)$, we use the notation $(V \indep U)_{Mod}$ to denote independence between variables $V$ and $U$ under the causal model ($Mod$). We further let $\mathbb{E}_{Mod}\left(V^{U=u}\right)$ be the expectation of variable $V^{U=u}$ under causal model (\textit{Mod}). We will mostly consider such expectations for (\textit{Mod}) set to either the true longitudinal causal model $(L)$ or the over-simplified model $(S)$ used for the analysis (see Section  \ref{Subsec:Notation_OverSimplifiedModel}). Because expectations and probabilities involving observable variables only will always be computed under the true longitudinal causal model ($L$), we simply write $\mathbb{E}(V)$ and $\mathbb{P}(V=v)$ for any observable  variable $V$ and any potential value $v$ of $V$, instead of $\mathbb{E}_L(V)$ and $\mathbb{P}_L(V=v)$, respectively.

In this framework, the causal effect of interest is that of the time-varying exposure $(X_t)_t$ on the outcome $Y$, and a key quantity in our work is therefore 
\begin{eqnarray}
ATE_L\left( \underline x_{t_0}^{t_3} ; \underline x_{t_1}^{t_2*}\right) = \mathbb{E}_L \left( Y^{\underline X_{t_0}^{t_3} = \underline x_{t_0}^{t_3} }   -  Y^{\underline X_{t_0}^{t_3} = \underline x_{t_0}^{t_3*} } \right), \label{eq:ATEL}
\end{eqnarray} for given profiles $\underline x_{t_0}^{t_3}$ and  $\underline x_{t_0}^{t_3*}$ for the exposure of interest, and given times $1 \leq t_0 \leq t_3 \leq T$.  This quantity is one measure of the total causal effect of the exposure from time $t_0$ to time $t_3$ on the outcome $Y$, under the longitudinal causal model $(L)$ \citep{G_Formula, VDW_Book}. In particular, total causal effects of the exposure over the full time interval may be considered ($t_0 = 1$, $t_3 = T$), but other total causal effects, including that of the exposure at a single time point ($t_0 = t_3$) can be considered as well under this longitudinal setting. 
Because the causal effect in Equation \eqref{eq:ATEL} generally depends on the particular profiles $\underline x_{t_0}^{t_3}$ and  $\underline x_{t_0}^{t_3*}$, averaged total effects can be defined as $$\sum_{\underline x_{t_0}^{t_3}}\sum_{\underline x_{t_0}^{t_3*}} ATE_L\left( \underline x_{t_0}^{t_3} ; \underline x_{t_0}^{t_3*}\right) \omega \left(\underline x_{t_0}^{t_3}, \underline x_{t_0}^{t_3*} \right), $$ for appropriate weights  $\omega(\underline x_{t_0}^{t_3},\underline x_{t_0}^{t_3*})$, and with the two sums over $\{0,1\} \times \dots \times \{0,1\} = \{0,1\}^{t_3 - t_0 + 1}$. For future use, we also introduce stratum-specific causal effects \citep{H_R}, with strata defined according to the levels of a possibly multivariate variable $U$
\begin{eqnarray}
ATE_{L_{\mid U = u}} \left( \underline x_{t_0}^{t_3} ; \underline x_{t_0}^{t_3*}\right)  :=\mathbb{E}_L\left( Y^{\underline X_{t_0}^{t_3} = \underline x_{t_0}^{t_3} }   -  Y^{\underline X_{t_0}^{t_3} = \underline x_{t_0}^{t_3*} } \mid U = u\right). \label{eq:stratum_sp_effect} 
\end{eqnarray}
Weighted averages of the form $\sum_{u}\sum_{\underline x_{t_0}^{t_3}}\sum_{\underline x_{t_0}^{t_3*}} ATE_{L_{\mid U = u}} \left( \underline x_{t_0}^{t_3} ; \underline x_{t_0}^{t_3*}\right) \omega \left(\underline x_{t_0}^{t_3},\underline x_{t_0}^{t_3*}, u \right) $ will also be considered for appropriate weights $\omega \left(\underline x_{t_0}^{t_3},\underline x_{t_0}^{t_3*}, u \right)$.

\iffalse
\begin{center}
\begin{tikzpicture}[scale=0.8, auto,swap]
\node[var] (Xm)at(4.9,1.25){$X_{1}$};
\node[var] (Xmm)at(6.6,1.25){$X_{2}$};
\node[var] (X)at(8.3,1.25){$\mathcall{X}$};
\node[var] (Wm)at(4.3,2.75){${W}_{1}$};
\node[var] (Wmm)at(6.0,2.75){$W_{2}$};
\node[var] (W)at(7.7,2.75){$\mathcall{W}$};
\node[var] (Y)at(9.3,2.2){$Y$};
\draw[edge] (Wm)--(Xm);
\draw[edge] (Wmm)--(Xmm);
\draw[edge] (Wm).. controls (7.85,3.55) ..(Y);
\draw[edge] (Xm).. controls (8.7,0.45) ..(Y);
\draw[edge] (Wmm).. controls (7.75,3.35) ..(Y);
\draw[edge] (Xmm).. controls (8.6,0.65) ..(Y);
\draw[edge] (Wmm)--(W);
\draw[edge] (Wm)--(Wmm);
\draw[edge] (Xm)--(Xmm);
\draw[edge] (Xm)--(Wmm);
\draw[edge] (Wm)--(Xmm);
\draw[edge] (Xmm)--(X);
\draw[edge] (X)--(Y);
\draw[edge] (W)--(Y);
\end{tikzpicture}
\end{center}
\fi

\subsection{The over-simplified causal model considered in practice}\label{Subsec:Notation_OverSimplifiedModel}

Often, analysts do not have access to, or simply ignore, the full exposures profiles ($\bar X_{T}, \bar Z_{T}$), and focus on  ``summaries'' $(\mathcall{X}, \mathcall{Z})$ of ($\bar X_{T}, \bar Z_{T}$) instead, with $\mathcall{X}= f_X(\bar X_{T})$ and $\mathcall{Z}=f_Z(\bar Z_{T})$, for two given deterministic functions $f_X$ and $f_Z$.  The most common case in practice is when $\mathcall{X}=X_{t}$ and $\mathcall{Z}=Z_{t}$, for some time $t \in \llbracket 1; T \rrbracket$, {\it e.g.}, the time of inclusion in the study. But $\mathcall{X}$ and $\mathcall{Z}$ can represent other summaries of ($\bar X_{T}, \bar Z_{T}$), such as $(i)$ cumulative exposure over some time-interval: $\mathcall{X} = \sum _{t=t_1} ^{t_2} X_t$ for some $1\leq t_1 \leq t_2 \leq T$; or $(ii)$  duration of exposure over a certain threshold $\mathcall{X} = \mathbb{1} \lbrace \sum _{t=t_1} ^{t_2} X_t \geq \tau \rbrace$ for some threshold $\tau \in \mathbb{R}$ \citep{ARNO, KUNZ, DeRubeis19, Yang19, Zheng18, Fan08, Platt10, Arnold19}. Moreover, $f_X$ and $f_Z$ do not have to share the same form; for example, we can have $\mathcall{Z}=Z_{t_1}$ and  $\mathcall{X} = \sum _{t=t_1} ^{t_2} X_t$. Lastly, some components of $\mathcall{Z}$ may be unobserved in practice.

When focusing on $\mathcall{X}$ and $\mathcall{Z}$, analysts generally ignore 
$\bar X_{T}$ and $\bar Z_{T}$ and work under an over-simplified causal model (\textit{S}), which only involves $\mathcall{X}$,  $\mathcall{Z}$ and $Y$. For example, if the true causal model is model $(L.1)$ of Figure \ref{fig:example_longitudinal_XMW} but only summaries $\mathcall{X}$ and $\mathcall{W}$ of $\bar X_{T}$ and $\bar W_{T}$ are available or considered, analysts may be tempted to work under the simplified causal models $(S.1)$ or $(S.2)$ of Figure \ref{fig:example_oversimplified_XMW}, depending on whether $(W_t)_{t \leq 1}$ is mainly regarded as a confounder or as a mediator. Consequently, they also generally consider $ATE_{S} \left(\mathcall{x} ; \mathcall{x}^{*}\right)   =\mathbb{E}_{S}\big( Y^{\mathcall{X} = \mathcall{x}} - Y^{\mathcall{X} = \mathcall{x}^{*}}\big)$, for any $\mathcall{x} \neq \mathcall{x}^{*}$, as the causal effect of interest. If there exists some observed $\mathcall{W} \subset \mathcall{Z}$ taking its values in $\Omega_{\mathcall{W}}$, such that $(Y^{\mathcall{X} = \mathcall{x}} \indep \mathcall{X} \mid \mathcall{W})_{S}$, as under model ($S.1$) of Figure \ref{fig:example_oversimplified_XMW}, and if some positivity condition holds \citep{rosenbaum1983}, analysts would identify, and then estimate, $ATE_{S} \left(\mathcall{x} ; \mathcall{x}^{*}\right)$ as 
$$\sum_{\mathcall{w} \in \Omega_{\mathcall{W}}} \left[ \mathbb{E}\left( Y \mid \mathcall{W} = \mathcall{w}, \mathcall{X} = \mathcall{x} \right) - \mathbb{E}\left( Y \mid \mathcall{W} = \mathcall{w}, \mathcall{X} = \mathcall{x}^{*} \right) \right]\times  \mathbb{P}(\mathcall{W}=\mathcall{w}).$$
Conversely, if $(Y^{\mathcall{X} = \mathcall{x}} \indep \mathcall{X})_{S}$, as under models ($S.2$) and ($S.3$) of Figure \ref{fig:example_oversimplified_XMW}, analysts would simply identify $ATE_{S} \left(\mathcall{x} ; \mathcall{x}^{*}\right)$ as $\mathbb{E}\left( Y \mid \mathcall{X} = \mathcall{x} \right) - \mathbb{E}\left( Y \mid  \mathcall{X} = \mathcall{x}^{*} \right)$.
We will thereafter denote by $\widetilde{ATE}_S(\mathcall{x} ; \mathcall{x}^{*})$ the observable quantity estimated in practice when working under the simplified causal model $(S)$.

\begin{center}
\begin{figure}
\begin{minipage}[c]{0.3\linewidth}
\begin{center}
\begin{tikzpicture}[scale=0.78, auto,swap]
\node[var] (X)at(6,0){$\mathcall{X}$};
\node[var] (Y)at(9.4,0){$Y$};
\node[var] (W)at(7.7,1){$\mathcall{W}$};
\draw[edge] (W)--(Y);
\draw[edge] (W)--(X);
\draw[edge] (X)--(Y);
\end{tikzpicture}
\end{center}
\end{minipage}\hfill
\begin{minipage}[c]{0.3\linewidth}
\begin{center}
\begin{tikzpicture}[scale=0.78, auto,swap]
\node[var] (X)at(6,0){$\mathcall{X}$};
\node[var] (Y)at(9.4,0){$Y$};
\node[var] (M)at(7.7,1){$\mathcall{M}$};
\draw[edge] (M)--(Y);
\draw[edge] (X)--(M);
\draw[edge] (X)--(Y);
\end{tikzpicture}
\end{center}
\end{minipage}\hfill
\begin{minipage}[c]{0.3\linewidth}
\begin{center}
\begin{tikzpicture}[scale=0.78, auto,swap]
\node[var] (X)at(6,0){$\mathcall{X}$};
\node[var] (Y)at(9.4,0){$Y$};
\draw[edge] (X)--(Y);
\end{tikzpicture}
\end{center}

\end{minipage}

\begin{minipage}[c]{0.3\linewidth}
\centering
($S.1$)
\end{minipage}\hfill
\begin{minipage}[c]{0.3\linewidth}
\centering
($S.2$)
\end{minipage}
\hfill
\begin{minipage}[c]{0.3\linewidth}
\centering
($S.3$)
\end{minipage}
    \caption{Examples of over-simplified causal models.}
    \label{fig:example_oversimplified_XMW}
\end{figure}
\end{center}

\section{Main result} \label{Sec:MainRes}

\subsection{Statement}
Because, the true causal model ($L$) involves the time-varying exposure of interest and time-varying auxiliary variables (see for example model ($L.1$) in Figure \ref{fig:example_longitudinal_XMW}), a natural question is whether - and how - this quantity $\widetilde{ATE}_{S} \left(\mathcall{x} ; \mathcall{x}^{*}\right)$  relates to causal effects of interest under the true longitudinal causal model $(L)$. Theorem \ref{Theo_SV_Weak} below presents a sufficient condition under which the quantity $\widetilde{ATE}_{S} \left(\mathcall{x} ; \mathcall{x}^{*}\right)$ estimated in practice can be written as a weighted average of (possibly stratum-specific) longitudinal total effects.

\begin{Theorem}\label{Theo_SV_Weak}
Assume that the observed $\mathcal X$ is a deterministic function of $\underline X_{t_1} ^{t_2}$ for some $1 \leq t_1 \leq t_2 \leq T$. 
For any possible values $\mathcall{x}$ and $\mathcall{x}^{*}$ of $\mathcal X$, if condition $(T.Cond)$ below holds
\begin{itemize}
\item[] (T.Cond) \quad There exists some $t_0 \in \llbracket 1, t_1 \rrbracket$, some $t_3 \in \llbracket t_2, T \rrbracket$ and some observed $\mathcall{W} \subset \mathcall{Z}$ taking its values in $\Omega_{\mathcall{W}}$, such that $(Y^{\mathcall{X} = \mathcall{x}} \indep \mathcall{X} \mid \mathcall{W})_{S}$, $0 < \mathbb{P}(\mathcall{X} = \mathcall{x} \mid \mathcall{W} = \mathcall{w}) <1$, $0 < \mathbb{P}(\mathcall{X} = \mathcall{x}^{*} \mid \mathcall{W} = \mathcall{w})<1$, for all $\mathcall{w}$ such that $0< \mathbb{P}(\mathcall{W} = \mathcall{w})$, and $(Y^{\underline X_{t_0}^{t_3} = \underline x_{t_0}^{t_3}} \indep \underline X_{t_0}^{t_3} \mid \mathcall{W})_{L}$,
\end{itemize}
then the quantity estimated in practice
\begin{eqnarray*}
\widetilde{ATE}_{S}  \left(\mathcall{x} ; \mathcall{x}^{*}\right) & = & \sum_{\mathcall{w} \in \Omega_{\mathcall{W}}} \left[ \mathbb{E}\left( Y \mid \mathcall{W} = \mathcall{w}, \mathcall{X} = \mathcall{x} \right) - \mathbb{E}\left( Y \mid \mathcall{W} = \mathcall{w}, \mathcall{X} = \mathcall{x}^{*} \right) \right]\\[-0.3cm]
&&  \hspace{1.2cm} \times  \mathbb{P}(\mathcall{W}=\mathcall{w}),
\end{eqnarray*}
equals
\begin{eqnarray}
&& \hspace{-0.2cm} \sum_{\mathcall{w}  \in \Omega_{\mathcall{W}}} \sum_{\substack{ \underline x_{t_0}^{t_3} \in \lbrace 0,1 \rbrace^{t_3 - t_0 + 1} \\  \underline x_{t_0}^{t_3*} \in \lbrace 0,1 \rbrace^{t_3 - t_0 + 1}}} \lbrace ATE_{L_{\mid \mathcall{W} = \mathcall{w}}}\left( \underline x_{t_0}^{t_3} ; \underline x_{t_0}^{t_3*} \right) \times \mathbb{P}(\underline X_{t_0}^{t_3} = \underline x_{t_0}^{t_3} \mid \mathcall{X} = \mathcall{x}, \mathcall{W} = \mathcall{w})  \nonumber   \\[-1cm]
&& \hspace{6.85cm} \times \mathbb{P}(\underline X_{t_0}^{t_3} = \underline x_{t_0}^{t_3*} \mid \mathcall{X} = \mathcall{x}^{*}, \mathcall{W} = \mathcall{w}) \nonumber \\
&& \hspace{6.85cm} \times \mathbb{P}(\mathcall{W} = \mathcall{w})\rbrace. \label{Eq:SV_Sufficient_cond}
\end{eqnarray}
In particular, if condition $(T.Uncond)$ below holds
\begin{itemize}
\item[] $(T.Uncond)$ There exists some $t_0 \in \llbracket 1, t_1 \rrbracket$ and some $t_3 \in \llbracket t_2, T \rrbracket$ such that \quad $(Y^{\mathcall{X} = \mathcall{x}} \indep \mathcall{X})_{S}$, $0 < \mathbb{P}(\mathcall{X} = \mathcall{x})<1$, $0 < \mathbb{P}(\mathcall{X} = \mathcall{x}^{*})<1$, and $(Y^{\underline X_{t_0}^{t_3} = \underline x_{t_0}^{t_3}} \indep \underline X_{t_0}^{t_3})_{L}$
\end{itemize}
then
\begin{eqnarray} 
\hspace{-0.2cm} \widetilde{ATE}_{S}(\mathcall{x} ; \mathcall{x}^*) & = &  \mathbb{E}\left( Y \mid\mathcall{X} = \mathcall{x} \right) - \mathbb{E}\left( Y \mid \mathcall{X} = \mathcall{x}^{*} \right),  \nonumber \\
& = & \sum_{\substack{\underline x_{t_0}^{t_3} \in \lbrace 0,1 \rbrace^{t_3 - t_0 + 1} \\  \underline x_{t_0}^{t_3*} \in \lbrace 0,1 \rbrace^{t_3 - t_0 + 1}}} \lbrace ATE_{L}\left( \underline x_{t_0}^{t_3} ; \underline x^{t_3*}_{t_0}\right) \times \mathbb{P}(\underline X_{t_0}^{t_3}  = \underline x_{t_0}^{t_3}  \mid \mathcall{X} = \mathcall{x})  \nonumber   \\[-0.9cm]
&& \hspace{5.2cm} \times \mathbb{P}(\underline X_{t_0}^{t_3} = \underline x_{t_0}^{t_3*}  \mid \mathcall{X} = \mathcall{x}^{*}) \rbrace.\label{Eq:SV_Sufficient_uncond}
\end{eqnarray}
\end{Theorem}

The proof of Theorem \ref{Theo_SV_Weak} is given in Appendix \ref{Proof:Theo_SV_Weak}. Theorem \ref{Theo_SV_Weak} states that when $\mathcall X$ depends on the exposure profile from time $t_1$ to time $t_2$ and there exists a possibly empty set of observed variables that satisfies (\textit{i}) the ignorability condition for the  exposure profile on a possibly wider time-window and the outcome under the true longitudinal model, and (\textit{ii}) the ignorability condition for $\mathcall{X}$ and the outcome under the over-simplified causal model, then the quantity estimated in practice expresses as a weighted average of possibly stratum-specific longitudinal total effects. Through the inspection of a few simple examples, we show in Section \ref{Sec:CondVV} that conditions $(T.Uncond)$ and $(T.Cond)$ are very restrictive, which confirms that the result of Theorem \ref{Theo_SV_Weak} is rarely valid in practice and that caution is usually required when interpreting the quantity estimated in practice. Indeed, even if conditions of Theorem \ref{Theo_SV_Weak} are sufficient conditions only, the quantity estimated in practice can generally not be expressed as the weighted average of any longitudinal effects of interest when they are not satisfied; see the Supplementary Material 1 and 2 for more details. Also, we show in Section \ref{Sec:VVImplications} that even when the conditions of Theorem \ref{Theo_SV_Weak} are satisfied, the interpretation of the weighted averages, hence that of the quantity estimated in practice, is sometimes not straightforward.

\subsection{About the conditions of Theorem \ref{Theo_SV_Weak}}\label{Sec:CondVV}

First, the presence of times $t_0 \in \llbracket 1, t_1 \rrbracket$ and $t_3 \in \llbracket t_2, T \rrbracket$ in conditions ($T.Cond$) and ($T.Uncond$) allows Theorem \ref{Theo_SV_Weak} to cover more general configurations, e.g., $(i)$ when $\mathcall{X}$ and $\mathcall{Z}$ are summaries of $\bar X_{T}$ and $\bar Z_{T}$ over different sub-intervals of $\llbracket 1, T \rrbracket$; or $(ii)$, when $\mathcall X = X_{t_2}$ and $\mathcall Z = Z_{t_2}$, for some time $t_2 \in \llbracket 1;T \rrbracket$. Specifically, condition ($T.Uncond$) of Theorem is satisfied with $t_0 = 1$ and $t_3 = t_2$ if the true causal model and simplified model are models $(L.0)$ and $(S.0)$ of Figure \ref{Fig:Example_Instantaneous}, respectively, and $\mathcall{X} = X_{t_2}$.

\begin{center}
\begin{figure}
\begin{minipage}[c]{0.5\linewidth}
\begin{center}
\begin{tikzpicture}[scale=0.78, auto,swap]
\node[var] (X)at(4,0){$\bar X_{t_2-1} $};
\node[var] (Xt)at(6,0){$X_{t_2}$};
\node[var] (Xtt)at(8,0){$\underline X_{t_2+1} ^T$};
\draw[edge] (X)--(Xt);
\draw[edge] (Xt)--(Xtt);
\node[var] (Y)at(10,0){$Y$};
\draw[edge] (X).. controls (7.5,-1.15) ..(Y);
\draw[edge] (X).. controls (6,0.75) ..(Xtt);
\draw[edge] (Xt).. controls (8,-0.75) ..(Y);
\draw[edge] (Xtt)--(Y);
\end{tikzpicture}
\end{center}
\end{minipage}\hfill
\begin{minipage}[c]{0.5\linewidth}
\begin{center}
\begin{tikzpicture}[scale=1, auto,swap]
\node[var] (X)at(2,0){$X_{t_2}$};
\node[var] (Y)at(4,0){$Y$};
\draw[edge] (X)--(Y);
\end{tikzpicture}
\end{center}
\end{minipage}

\vspace{0.2cm}

\begin{minipage}[c]{0.5\linewidth}
\begin{center}
(\textit{L.0})
\end{center}
\end{minipage}\hfill
\begin{minipage}[c]{0.5\linewidth}
\begin{center}
(\textit{S.0})
\end{center}
\end{minipage}
 \caption{(\textit{L.0}) Example of longitudinal model with a time-varying exposure of interest $(X_t)_{t \geq 1}$, in the absence of confounding. (\textit{S.0}) Over-simplified causal model associated with the longitudinal model (\textit{L.0}), when data on the exposure is considered at time $t_2$ only.}
    \label{Fig:Example_Instantaneous}
\end{figure}
\end{center}

Second, conditions ($T.Cond$) and ($T.Uncond$) of Theorem \ref{Theo_SV_Weak} are not testable, but the backdoor criterion can be applied to the DAGs corresponding to models $(S)$ and $(L)$, respectively, to check their validity (assuming these DAGs are well specified). For the true causal model $(L)$, the ``augmented'' DAG where summary variables $\mathcall{X}$ and $\mathcall{W}$ are explicitly represented should be considered. For example, under the longitudinal causal model $(L.1)$ of Figure \ref{fig:example_longitudinal_XMW}, and assuming that the observed summary variables $\mathcall{X}$ and $\mathcall{W}$ are functions of the full exposure profiles $\bar X_{T}$ and $\bar W_{T}$ (that is, $t_1=1$ and $t_2=T$), possible augmented DAGs are presented in Figure \ref{fig:AugmentedVV}, depending on whether these summary variables capture the whole effect of $\bar X_{T}$ and $\bar W_{T}$ on $Y$ or not. In either case, neither $\mathcall{W}$ nor the empty set satisfies the backdoor criterion relative to $\bar X_{T}$ and $Y$ in these augmented DAGs, so neither ($T.Cond$) nor ($T.Uncond$) is satisfied, and Theorem  \ref{Theo_SV_Weak} does not apply.

\begin{center}
\begin{figure}
\begin{minipage}[c]{0.5\linewidth}

\begin{tikzpicture}[scale=0.8, auto,swap]
\node[var] (W2)at(2,0){$W_1$};
\node[var] (W3)at(4,0){$W_2$};
\node[var] (Wdot)at(5.5,0){$\dots$};
\node[var] (Wt)at(7,0){$W_{T}$};
\node[var] (W)at(8.5,0){$\mathcall{W}$};
\node[var] (X2)at(2,-1.5){$X_1$};
\node[var] (X3)at(4,-1.5){$X_2$};
\node[var] (Xdot)at(5.5,-1.5){$\dots$};
\node[var] (Xt)at(7,-1.5){$X_{T}$};
\node[var] (X)at(8.5,-1.5){$\mathcall{X}$};
\node[var] (Y)at(9.75,-0.75){$Y$};

\draw[edge] (X2)--(X3);
\draw[edge] (X2)--(W3);
\draw[edge] (X2).. controls (3.5, -0.85)..(Wt);
\draw[edge] (X3)--(Wt);

\draw[edge] (W2)--(W3);
\draw[edge] (W2)--(X2);
\draw[edge] (W2)--(X3);
\draw[edge] (W2).. controls (3.5, -0.75)..(Xt);
\draw[edge] (W3)--(Xt);

\draw[edge] (W3)--(X3);
\draw[edge] (W3)--(Wdot);
\draw[edge] (Wdot)--(Wt);
\draw[edge] (Wt)--(W);
\draw[edge] (Wt)--(Xt);

\draw[edge] (X3)--(Xdot);
\draw[edge] (Xdot)--(Xt);
\draw[edge] (Xt)--(X);

\draw[edge] (X)--(Y);
\draw[edge] (W)--(Y);

\draw[edge] (W2).. controls (5.1,0.75) ..(Wt);
\draw[edge] (W2).. controls (5,1.15) ..(W);

\draw[edge] (W3).. controls (5.1,0.5) ..(Wt);
\draw[edge] (W3).. controls (6.2,1.15) ..(W);

\draw[edge] (X2).. controls (5.1,-2.2) ..(Xt);
\draw[edge] (X2).. controls (5,-2.5) ..(X);
\draw[edge] (X2).. controls (5,-3.5) and (8.5,-3.5)..(Y);

\draw[edge] (X3).. controls (5.1,-1.9) ..(Xt);
\draw[edge] (X3).. controls (6.2,-2.5) ..(X);

\draw[edge] (X3).. controls (6.2,-3.1) and (7.9,-3.1)..(Y);

\draw[edge] (Xt).. controls (7.1,-2.5) and (7.9,-2.5)  ..(Y);

\draw[edge] (W2).. controls (5,2.15) and (8.5,2.15)..(Y);
\draw[edge] (W3).. controls (6.2,1.75) and (7.9,1.75)..(Y);
\draw[edge] (Wt).. controls (7.1,1.15) and (7.9,1.15)  ..(Y);

\end{tikzpicture}
\end{minipage}\hfill
\begin{minipage}[c]{0.5\linewidth}

\begin{tikzpicture}[scale=0.8, auto,swap]
\node[var] (W2)at(2,0){$W_1$};
\node[var] (W3)at(4,0){$W_2$};
\node[var] (Wdot)at(5.5,0){$\dots$};
\node[var] (Wt)at(7,0){$W_{T}$};
\node[var] (W)at(8.5,0){$\mathcall{W}$};
\node[var] (X2)at(2,-1.5){$X_1$};
\node[var] (X3)at(4,-1.5){$X_2$};
\node[var] (Xdot)at(5.5,-1.5){$\dots$};
\node[var] (Xt)at(7,-1.5){$X_{T}$};
\node[var] (X)at(8.5,-1.5){$\mathcall{X}$};
\node[var] (Y)at(9.75,-0.75){$Y$};

\draw[edge] (X2)--(X3);
\draw[edge] (X2)--(W3);
\draw[edge] (X2).. controls (3.5, -0.85)..(Wt);
\draw[edge] (X3)--(Wt);

\draw[edge] (W2)--(W3);
\draw[edge] (W2)--(X2);
\draw[edge] (W2)--(X3);
\draw[edge] (W2).. controls (3.5, -0.75)..(Xt);
\draw[edge] (W3)--(Xt);

\draw[edge] (W3)--(X3);
\draw[edge] (W3)--(Wdot);
\draw[edge] (Wdot)--(Wt);
\draw[edge] (Wt)--(W);
\draw[edge] (Wt)--(Xt);

\draw[edge] (X3)--(Xdot);
\draw[edge] (Xdot)--(Xt);
\draw[edge] (Xt)--(X);

\draw[edge] (X)--(Y);
\draw[edge] (W)--(Y);

\draw[edge] (W2).. controls (5.1,0.75) ..(Wt);
\draw[edge] (W2).. controls (5,1.15) ..(W);

\draw[edge] (W3).. controls (5.1,0.5) ..(Wt);
\draw[edge] (W3).. controls (6.2,1.15) ..(W);

\draw[edge] (X2).. controls (5.1,-2.2) ..(Xt);
\draw[edge] (X2).. controls (5,-2.5) ..(X);

\draw[edge] (X3).. controls (5.1,-1.9) ..(Xt);
\draw[edge] (X3).. controls (6.2,-2.5) ..(X);

\end{tikzpicture}
\end{minipage}

\begin{minipage}[c]{0.5\linewidth}
\centering
(a)
\end{minipage} \hfill
\begin{minipage}[c]{0.5\linewidth}
\centering
(b)
\end{minipage} 
\caption{Possible (augmented) DAGs corresponding to longitudinal causal model associated with model (\textit{L}.1) given in Figure \ref{fig:example_longitudinal_XMW}, depending on whether summaries $\mathcall W$ and $\mathcall{X}$ do not (a) or do (b) catpure the whole effect of $\bar W_T$ and $\bar X_T$ on $Y$.}
\label{fig:AugmentedVV}
\end{figure}
\end{center}

Conversely, if the summary variables $\mathcall{X}$ and $\mathcall{W}$ capture the whole effect of $\bar X_{T}$ and $\bar W_{T}$, conditions of Theorem \ref{Theo_SV_Weak} are satisfied in either ($i$) the pure confounder setting of model ($L.2$) of Figure \ref{Fig:ATE_SV_example2_configuration} (if, in addition, the over-simplified model is model ($S.1$) of Figure \ref{fig:example_oversimplified_XMW}) or ($ii$) the pure mediator setting of model ($L.3$) of Figure \ref{Fig:ATE_SV_example2_configuration} (if the over-simplified model is model ($S.2$), or ($S.3$), of Figure \ref{fig:example_oversimplified_XMW}). When working under model ($S.1$) while the true causal model is ($L.2$), condition ($T.Cond$) holds and Theorem \ref{Theo_SV_Weak} states that  $\widetilde{ATE}_{S}(\mathcall{x}; \mathcall{x}^*)$ is the weighted average of the longitudinal total effects given in Equation \eqref{Eq:SV_Sufficient_cond}. When working under model ($S.2$) (or ($S.3$)) while the true causal model is ($L.3$), condition ($T.Uncond$) holds and Theorem \ref{Theo_SV_Weak} states that  $\widetilde{ATE}_{S}(\mathcall{x}; \mathcall{x}^*)$ is the weighted average of the longitudinal total effects given in Equation \eqref{Eq:SV_Sufficient_uncond}. Interestingly, the conditions of Theorem \ref{Theo_SV_Weak} are not satisfied when the true causal model is model ($L.4$) of Figure \ref{Fig:ATE_SV_example3_configuration}, where both a pure time-varying confounder and a pure time-varying mediator that is affected by the confounder are present. Indeed, again, neither ($T.Cond$) nor ($T.Uncond$) is satisfied, and Theorem  \ref{Theo_SV_Weak} does not apply. This is in sharp contrast with the setting of model ($L.2$), where only a time-varying pure confounder, and no time-varying pure mediator, is present, and in which case Theorem \ref{Theo_SV_Weak} applies. In other words, although generally overlooked when the focus is on total effects, the presence of a time-varying mediator affected by a pure confounder precludes the validity of Theorem \ref{Theo_SV_Weak}.

\begin{center}
\begin{figure}
\begin{minipage}[c]{0.48\linewidth}
\begin{center}
\begin{tikzpicture}[scale=0.71, auto,swap]
\node[var] (W1)at(0.89,0){$W_1$};
\node[var] (W2)at(2.6,0){$\dots$};
\node[var] (W3)at(4,0){$W_{t}$};
\node[var] (Wdot)at(5.5,0){$\dots$};
\node[var] (Wt)at(7,0){$W_{T}$};
\node[var] (W)at(8.5,0){$\mathcall{W}$};
\node[var] (X1)at(0.89,-1.5){$X_1$};
\node[var] (X2)at(2.6,-1.5){$\dots$};
\node[var] (X3)at(4,-1.5){$X_{t}$};
\node[var] (Xdot)at(5.5,-1.5){$\dots$};
\node[var] (Xt)at(7,-1.5){$X_{T}$};
\node[var] (X)at(8.5,-1.5){$\mathcall{X}$};
\node[var] (Y)at(9.75,-0.75){$Y$};

\draw[edge] (X1)--(X2);
\draw[edge] (X2)--(X3);

\draw[edge] (W1)--(W2);
\draw[edge] (W2)--(W3);

\draw[edge] (W1)--(X1);
\draw[edge] (W1)--(X3);
\draw[edge] (W1).. controls (3.5, -0.75)..(Xt);
\draw[edge] (W3)--(Xt);

\draw[edge] (W3)--(X3);
\draw[edge] (W3)--(Wdot);
\draw[edge] (Wdot)--(Wt);
\draw[edge] (Wt)--(W);
\draw[edge] (Wt)--(Xt);

\draw[edge] (X3)--(Xdot);
\draw[edge] (Xdot)--(Xt);
\draw[edge] (Xt)--(X);

\draw[edge] (X)--(Y);
\draw[edge] (W)--(Y);

\draw[edge] (W1).. controls (4,1.15) ..(Wt);
\draw[edge] (W1).. controls (3,0.65) ..(W3);
\draw[edge] (W1).. controls (5.3,1.75) ..(W);

\draw[edge] (W3).. controls (5.1,0.5) ..(Wt);
\draw[edge] (W3).. controls (6.2,1.15) ..(W);

\draw[edge] (X1).. controls (4,-2.5) ..(Xt);
\draw[edge] (X1).. controls (3,-2.1) ..(X3);
\draw[edge] (X1).. controls (5.2,-3.2) ..(X);

\draw[edge] (X3).. controls (5.1,-1.9) ..(Xt);
\draw[edge] (X3).. controls (6.2,-2.5) ..(X);
\end{tikzpicture}
\end{center}
\end{minipage}\hfill
\begin{minipage}[c]{0.48\linewidth}
\begin{center}
\begin{tikzpicture}[scale=0.71, auto,swap]
\node[var] (W1)at(0.89,0){$M_1$};
\node[var] (W2)at(2.6,0){$\dots$};
\node[var] (W3)at(4,0){$M_{t}$};
\node[var] (Wdot)at(5.5,0){$\dots$};
\node[var] (Wt)at(7,0){$M_{T}$};
\node[var] (W)at(8.5,0){$\mathcall{M}$};
\node[var] (X1)at(0.89,-1.5){$X_1$};
\node[var] (X2)at(2.6,-1.5){$\dots$};
\node[var] (X3)at(4,-1.5){$X_{t}$};
\node[var] (Xdot)at(5.5,-1.5){$\dots$};
\node[var] (Xt)at(7,-1.5){$X_{T}$};
\node[var] (X)at(8.5,-1.5){$\mathcall{X}$};
\node[var] (Y)at(9.75,-0.75){$Y$};

\draw[edge] (X1)--(X2);
\draw[edge] (X2)--(X3);

\draw[edge] (W1)--(W2);
\draw[edge] (W2)--(W3);

\draw[edge] (X1)--(W1);
\draw[edge] (X1)--(W3);
\draw[edge] (X1).. controls (3.5, -0.75)..(Wt);
\draw[edge] (X3)--(Wt);

\draw[edge] (X3)--(W3);
\draw[edge] (W3)--(Wdot);
\draw[edge] (Wdot)--(Wt);
\draw[edge] (Wt)--(W);
\draw[edge] (Xt)--(Wt);

\draw[edge] (X3)--(Xdot);
\draw[edge] (Xdot)--(Xt);
\draw[edge] (Xt)--(X);

\draw[edge] (X)--(Y);
\draw[edge] (W)--(Y);

\draw[edge] (W1).. controls (4,1.15) ..(Wt);
\draw[edge] (W1).. controls (3,0.65) ..(W3);
\draw[edge] (W1).. controls (5.3,1.75) ..(W);

\draw[edge] (W3).. controls (5.1,0.5) ..(Wt);
\draw[edge] (W3).. controls (6.2,1.15) ..(W);

\draw[edge] (X1).. controls (4,-2.5) ..(Xt);
\draw[edge] (X1).. controls (3,-2.1) ..(X3);
\draw[edge] (X1).. controls (5.2,-3.2) ..(X);

\draw[edge] (X3).. controls (5.1,-1.9) ..(Xt);
\draw[edge] (X3).. controls (6.2,-2.5) ..(X);

\end{tikzpicture}
\end{center}
\end{minipage}\hfill

%\begin{minipage}[c]{0.02\linewidth}
%\end{minipage}

%\vspace{0.1cm}

\begin{minipage}[c]{0.48\linewidth}
\begin{center}
(\textit{L.2})
\end{center}
\end{minipage}\hfill
\begin{minipage}[c]{0.48\linewidth}
\begin{center}
(\textit{L.3})
\end{center}
\end{minipage}

    \caption{(\textit{L.2}) Example of a longitudinal causal model with a time-varying exposure of interest $(X_t)_{t\geq 1}$ and a time-varying pure confounder $(W_t)_{t\geq 1}$. (\textit{L.3}) Example of a longitudinal causal model with a time-varying exposure of interest $(X_t)_{t\geq 1}$ and a time-varying pure mediator $(M_t)_{t\geq 1}$. In both cases, exposures of interest, confounders and mediators are assumed to affect the outcome through the observed summary variables only.}\label{Fig:ATE_SV_example2_configuration}
\end{figure}
\end{center}

\begin{center}
\begin{figure}
\begin{minipage}[c]{0.48\linewidth}
\begin{center}
\begin{tikzpicture}[scale=0.85, auto,swap]
\node[var] (Xmm)at(6.5,0){$\bar X_{T}$};
\node[var] (Mmm)at(7.2,1.25){$\bar M_{T}$};
\node[var] (X)at(8.5,0){$\mathcall{X}$};
\node[var] (M)at(8.8,1.25){$\mathcall{M}$};
\node[var] (Wmm)at(5.8,2.5){$\bar W_{T}$};
\node[var] (W)at(8.2,2.5){$\mathcall{W}$};
\node[var] (Y)at(10,0.625){$Y$};
\draw[edge] (Wmm)--(W);
\draw[edge] (Xmm)--(X);
\draw[edge] (Mmm)--(M);
\draw[edge] (Xmm)--(Mmm);
\draw[edge] (Xmm)--(Mmm);
\draw[edge] (Wmm)--(Xmm);
\draw[edge] (Wmm)--(Mmm);
\draw[edge] (W)--(Y);
\draw[edge] (M)--(Y);
\draw[edge] (X)--(Y);
\end{tikzpicture}
\end{center}
\end{minipage}\hfill
\begin{minipage}[c]{0.48\linewidth}
\begin{center}
\begin{tikzpicture}[scale=0.92, auto,swap]
\node[var] (Y)at(4,-0.5){$Y$};
\node[var] (X)at(2,0){$\mathcall{X}$};
\node[var] (M)at(3.2,0.9){$\mathcall{M}$};
\node[var] (W)at(2.5,1.8){$\mathcall{W}$};
\draw[edge] (X)--(Y);
\draw[edge] (X)--(M);
\draw[edge] (M)--(Y);
\draw[edge] (W).. controls (3.9,1.4) ..(Y);
\draw[edge] (W)--(X);
\draw[edge] (W)--(M);
\end{tikzpicture}
\end{center}
\end{minipage}

\vspace{0.2cm}

\begin{minipage}[c]{0.48\linewidth}
\begin{center}
(\textit{L.4})
\end{center}
\end{minipage}\hfill
\begin{minipage}[c]{0.48\linewidth}
\begin{center}
(\textit{S.4})
\end{center}
\end{minipage}
    \caption{(\textit{L.4}) Compact representation of a longitudinal model with a time-varying exposure of interest $(X_t)_{t \geq 1}$, a time-varying pure mediator $(M_t)_{t \geq 1}$ and a time-varying pure confounder $(W_t)_{t \geq 1}$. These three processes affect the outcome $Y$ only through the observed summaries $\mathcall X$, $\mathcall{M}$ and $\mathcall{W}$, respectively, which are assumed to be deterministic functions of $\bar X_T$, $\bar M_T$ and $\bar W_T$, respectively. (\textit{S.4}) Over-simplified causal model associated with the longitudinal causal model given in Figure \ref{Fig:ATE_SV_example3_configuration} (\textit{L.4}).}
    \label{Fig:ATE_SV_example3_configuration}
\end{figure}
\end{center}

Another fundamental remark is that when $t_2 < T$, the conditions of Theorem \ref{Theo_SV_Weak} are generally not fulfilled, even under the simple case where only a pure confounder is present. For illustration, consider the simple model ($L.5$) of Figure \ref{fig:PureConfVV}. Because $\bar W_{t_2}$ affects not only $\bar X_{t_2}$ but also $\underline X_{t_2 + 1}^T$, there is no $t_0 \in \llbracket 1, t_1 \rrbracket$ and $t_3 \in \llbracket t_2, T \rrbracket$ such that $\mathcall W$ blocks all the backdoor paths between $Y$ and $\underline X_{t_0}^{t_3}$, and Theorem \ref{Theo_SV_Weak} is not satisfied (except under very particular settings under which the whole effect of $\bar W_{t_2}$ is captured by $\mathcall{X}$ and $\mathcall{W}$). Model ($L.5$) can be seen as a special case of model ($L.conf.gen$) of Figure \ref{fig:PureConfVV}, where summary variables do not capture the whole effect of $\bar W_{T}$ and $\bar X_{T}$, and where Theorem \ref{Theo_SV_Weak} does generally not apply since $\mathcall{W}$ does not block the backdoor path $Y \leftarrow \bar W_T \rightarrow \bar X_T$.

\begin{center}
\begin{figure}
\begin{minipage}[c]{0.54\linewidth}
\begin{center}
\begin{tikzpicture}[scale=0.75, auto,swap]
\node[var] (W1)at(0.8,0){$W_1$};
\node[var] (W2)at(2.6,0){$\dots$};
\node[var] (W3)at(4,0){$W_{t_2}$};
\node[var] (Wdot)at(5.5,0){$\dots$};
\node[var] (Wt)at(7,0){$W_{T}$};
\node[var] (W)at(5,1.5){$\mathcall{W}$};
\node[var] (X1)at(0.8,-1.5){$X_1$};
\node[var] (X2)at(2.6,-1.5){$\dots$};
\node[var] (X3)at(4,-1.5){$X_{t_2}$};
\node[var] (Xdot)at(5.5,-1.5){$\dots$};
\node[var] (Xt)at(7,-1.5){$X_{T}$};
\node[var] (X)at(5,-3){$\mathcall{X}$};
\node[var] (Y)at(9.75,-0.75){$Y$};

\draw[edge] (X1)--(X2);
\draw[edge] (X2)--(X3);

\draw[edge] (W1)--(W2);
\draw[edge] (W2)--(W3);

\draw[edge] (W1)--(X1);
\draw[edge] (W1)--(X3);
\draw[edge] (W1).. controls (3.5, -0.75)..(Xt);
\draw[edge] (W3)--(Xt);

\draw[edge] (W3)--(X3);
\draw[edge] (W3)--(Wdot);
\draw[edge] (Wdot)--(Wt);
\draw[edge] (Wt)--(Xt);

\draw[edge] (X3)--(Xdot);
\draw[edge] (Xdot)--(Xt);

\draw[edge] (X).. controls (8,-3.0) ..(Y);
\draw[edge] (W).. controls (8,1.5) ..(Y);

\draw[edge] (W1).. controls (4,1.15) ..(Wt);
\draw[edge] (W1).. controls (3,0.65) ..(W3);
\draw[edge] (W1).. controls (4, 2) ..(W);

\draw[edge] (W3).. controls (5.1,0.5) ..(Wt);
\draw[edge] (W3)--(W);

\draw[edge] (X1).. controls (4,-2.5) ..(Xt);
\draw[edge] (X1).. controls (3,-2.1) ..(X3);
\draw[edge] (X1).. controls (4,-3.5) ..(X);

\draw[edge] (X3).. controls (5.1,-1.9) ..(Xt);
\draw[edge] (X3)--(X);

\draw[edge] (Xt)--(Y);
\draw[edge] (Wt)--(Y);

\end{tikzpicture}
\end{center}
\end{minipage}\hfill
\begin{minipage}[c]{0.4\linewidth}

\begin{center}
\begin{tikzpicture}[scale=0.75, auto,swap]
\node[var] (W1)at(0.8,0){$\bar W_T$};
\node[var] (W)at(2.5,1.5){$\mathcall{W}$};
\node[var] (X1)at(0.8,-1.5){$\bar X_T$};
\node[var] (X)at(2.5,-3){$\mathcall{X}$};
\node[var] (Y)at(5,-0.75){$Y$};
\draw[edge] (W1)--(X1);
\draw[edge] (X)--(Y);
\draw[edge] (W)--(Y);

\draw[edge] (W1)--(W);
\draw[edge] (X1)--(X);

\draw[edge] (W1)--(Y);
\draw[edge] (X1)--(Y);

\end{tikzpicture}
\end{center}

\end{minipage}

\begin{minipage}[c]{0.54\linewidth}
\begin{center}
(\textit{L.5})
\end{center}
\end{minipage}\hfill
\begin{minipage}[c]{0.4\linewidth}
\begin{center}
(\textit{L.conf.gen})
\end{center}
\end{minipage}

    \caption{(\textit{L.5}) Example of a longitudinal causal model with a time-varying exposure of interest $(X_t)_{t\geq 1}$ and a time-varying pure confounder $(W_t)_{t\geq 1}$. These two processes potentially affect the outcome $Y$ through some summaries $\mathcall X$ and $\mathcall{W}$, respectively, which are assumed to be deterministic functions of $\bar X_{T}$ and $\bar W_{T}$, respectively. (\textit{L.conf.gen}) Compact representation of a longitudinal causal model with a pure confounder only, but where summary variables do not capture the whole effect of $\bar X_{T}$ and $\bar W_{T}$.}\label{fig:PureConfVV}
  
\end{figure}
\end{center}

\subsection{On the interpretation of the weighted averages in Theorem \ref{Theo_SV_Weak} when its conditions are satisfied}\label{Sec:VVImplications}

We now turn our attention to the weighted averages in Equations \eqref{Eq:SV_Sufficient_cond} and \eqref{Eq:SV_Sufficient_uncond} when conditions of Theorem \ref{Theo_SV_Weak} are satisfied. The summary variable $\mathcal X$ can be seen as a ``compound treatment'', with distinct exposure profiles $\underline x_{t_0} ^{t_3}$ leading to ${\mathcall X =\mathcall x}$, for any possible value ${\mathcall x}$ of ${\mathcall X}$, corresponding to distinct versions of this compound treatment ${\mathcall X}$, or ${\mathcall x}$ \citep{VDW_H, VDW_H_2}; here times $t_0$ and $t_3$ are those for which Condition ($T.Cond$) or ($T.Uncond$) of Theorem \ref{Theo_SV_Weak} is satisfied. Adopting the same terminology as in \cite{VDW_H}, we will say that versions of treatment ${\mathcall X}$ are irrelevant, when all versions $\underline x_{t_0} ^{t_3}$ leading to ${\mathcall X =\mathcall x}$ share the same effect on the outcome. %, which is then simply $Y^{{\mathcall X}={\mathcall x}}$. 
More formally,  versions of treatment ${\mathcall X}$  are irrelevant when condition  $(Irrel_{\llbracket t_0, t_3 \rrbracket})$ below holds:
\begin{itemize}
\item[] $(Irrel_{\llbracket t_0, t_3 \rrbracket})$ \quad $Y^{\underline X_{t_0} ^{t_3}=\underline x_{t_0} ^{t_3}} = Y^{{\mathcall X}={\mathcall x}}$ for any $\underline x_{t_0} ^{t_3}$ such that $(\underline X_{t_0} ^{t_3}=\underline x_{t_0} ^{t_3}) \Rightarrow ({\mathcall X}={\mathcall x})$.
\end{itemize}
Because ${\mathcall X}$ is a  deterministic function of $\bar X_T$, direct interventions on ${\mathcall X}$ cannot be implemented in practice. As a result, $Y^{{\mathcall X}={\mathcall x}}$, although mathematically grounded, does not always have a clear meaning. When versions are irrelevant, $Y^{{\mathcall X}={\mathcall x}}$ does have a clear meaning as it equals $Y^{\underline X_{t_0} ^{t_3}=\underline x_{t_0} ^{t_3}}$ for any $\underline x_{t_0} ^{t_3}$ such that $\underline X_{t_0} ^{t_3}=\underline x_{t_0} ^{t_3}$ implies  ${\mathcall X}={\mathcall x}$. For example, in model (\textit{L.2}) of Figure \ref{Fig:ATE_SV_example2_configuration}, with $t_0 = 1$ and $t_3 = T$, we have $ATE_{L.2}(\underline x_{t_0} ^{t_3} ; {\underline x_{t_0} ^{t_3}}^{*}) = ATE_{L.2}({\mathcall x} ; {\mathcall x}^{*}) = \mathbb{E}_{L.2}\big( Y^{\mathcall{X} = \mathcall{x}} $ $- Y^{\mathcall{X} = \mathcall{x}^{*}}\big)$, for any ${\underline x_{t_0} ^{t_3}}$ and ${\underline x_{t_0} ^{t_3}}^{*}$ leading to $\mathcall{X} = \mathcall{x}$ and $\mathcall{X} = \mathcall{x}^{*}$, respectively. To recap, when $(Irrel_{\llbracket t_0, t_3 \rrbracket})$ holds, the interpretation of the weighted averages in Equations \eqref{Eq:SV_Sufficient_cond} and \eqref{Eq:SV_Sufficient_uncond} is straigthforward as each term in the weighted averages simply equals $\mathbb{E}_{L.3}\left( Y^{\mathcall{X} = \mathcall{x}} - Y^{\mathcall{X} = \mathcall{x}^{*}}\right)$ (in the case of Equation \eqref{Eq:SV_Sufficient_cond}) or $\mathbb{E}_{L.3}\left( Y^{\mathcall{X} = \mathcall{x}} - Y^{\mathcall{X} = \mathcall{x}^{*}} |\mathcall{W}=\mathcall{w}\right)$ (in the case of Equation \eqref{Eq:SV_Sufficient_uncond}). %As a result, $\mathbb{E}_{L.3}\left( Y^{\mathcall{X} = \mathcall{x}} - Y^{\mathcall{X} = \mathcall{x}^{*}}\right)$ is well-defined and constitutes a causal effect of interest. 

 When $(Irrel_{\llbracket t_0, t_3 \rrbracket})$ does not hold, versions are relevant and we can have $Y^{{\underline X_{t_0} ^{t_3}} = {\underline x_{t_0} ^{t_3}}} \neq Y^{{\underline X_{t_0} ^{t_3}} = {\underline x_{t_0} ^{t_3'}}}$ for two exposure profiles $ {\underline x_{t_0} ^{t_3}}$ and $ {\underline x_{t_0} ^{t_3'}}$ leading to the same value $\mathcall{x}$ for $\mathcal{X}$.  For example, consider model ($L.1$), and more precisely the scenario of Figure \ref{fig:AugmentedVV} (b) with $t_0 = 1$ and $t_3 = T$ (similar arguments hold under the setting of Figure \ref{fig:AugmentedVV} (a)). Condition $(Irrel_{\llbracket t_0, t_3 \rrbracket})$ does not hold since ${\underline X_{t_0} ^{t_3}}$ affects $Y$ not only through $\mathcall{X}$, but also through some components of ${\mathcall W}$. Indeed, we can have ${\mathcall W}^{{\underline X_{t_0} ^{t_3}} = {\underline x_{t_0} ^{t_3}}} \neq  {\mathcall W}^{{\underline X_{t_0} ^{t_3}} = {\underline x_{t_0} ^{t_3'}}}$, and, in turn $Y^{{\underline X_{t_0} ^{t_3}} = {\underline x_{t_0} ^{t_3}}} \neq Y^{{\underline X_{t_0} ^{t_3}} = {\underline x_{t_0} ^{t_3'}}}$,  for two exposure profiles ${\underline x_{t_0} ^{t_3}}$ and ${\underline x_{t_0} ^{t_3'}}$ leading to the same value $\mathcall{X} = \mathcall{x}$. As a result, when versions are relevant, we typically have $ATE_{L}(\underline x_{t_0} ^{t_3} ; {\underline x_{t_0} ^{t_3}}^{*}) \neq ATE_{L}(\underline x_{t_0} ^{t_3'} ; {\underline x_{t_0} ^{t_3}}'^{*})$, even if both $\underline x_{t_0} ^{t_3}$ and $\underline x_{t_0} ^{t_3'}$ lead to $\mathcall{X} = \mathcall{x}$ and both $\underline x_{t_0} ^{t_3 *}$ and $\underline x_{t_0} ^{t_3'^{*}}$ lead to $\mathcall{X} = \mathcall{x}^{*}$. To better appreciate the meaning of the weighted averages of Equations \eqref{Eq:SV_Sufficient_cond} and \eqref{Eq:SV_Sufficient_uncond} when  $(Irrel_{\llbracket t_0, t_3 \rrbracket})$ does not hold, we can rewrite these weighted averages as follows. For example, the weighted average of Equation \eqref{Eq:SV_Sufficient_uncond} writes
\begin{eqnarray}
&& \hspace{-0.7cm} \sum_{{\underline x_{t_0} ^{t_3}}} \sum_{{\underline x_{t_0} ^{t_3}}^{*}} \lbrace ATE_{L}({\underline x_{t_0} ^{t_3}} ; {\underline x_{t_0} ^{t_3}}^{*}) \times \mathbb{P}({\underline X_{t_0} ^{t_3}} = {\underline x_{t_0} ^{t_3}} \mid \mathcall{X} = \mathcall{x}) \times \mathbb{P}({\underline X_{t_0} ^{t_3}} = {\underline x_{t_0} ^{t_3}}^{*} \mid \mathcall{X} = \mathcall{x}^{*})\rbrace\nonumber \\
&& \hspace{-0.3cm}  = \sum_{{\underline x_{t_0} ^{t_3}}} \lbrace \mathbb{E}_{L} \left(Y^{{\underline X_{t_0} ^{t_3} = \underline x_{t_0} ^{t_3}}} \right) \times \mathbb{P}({\underline X_{t_0} ^{t_3}} = {\underline x_{t_0} ^{t_3}} \mid \mathcall{X} = \mathcall{x})\rbrace  \nonumber \\
&& -\sum_{{\underline x_{t_0} ^{t_3}}^{*}} \lbrace\mathbb{E}_{L} \Big(Y^{{\underline X_{t_0} ^{t_3}}^{*}={\underline x_{t_0} ^{t_3}}^{*}} \Big) \times \mathbb{P}( {\underline X_{t_0} ^{t_3}} = {\underline x_{t_0} ^{t_3}}^* \mid \mathcall{X} = \mathcall{x}^*)\rbrace.\label{Average} 
\end{eqnarray}
From Equation \eqref{Average}, it follows that the weighted average of Equation \eqref{Eq:SV_Sufficient_uncond} represents the difference between the expectation of the outcome in the following two counterfactual populations. In the first one, a proportion $\mathbb{P}({\underline X_{t_0} ^{t_3}} = {\underline x_{t_0} ^{t_3}} \mid \mathcall{X} = \mathcall{x})$ of the individuals undergoes the intervention $do({\underline X_{t_0} ^{t_3}} = {\underline x_{t_0} ^{t_3}})$, for any profile ${\underline x_{t_0} ^{t_3}} $ leading to ${\mathcall X} = {\mathcall x}$. This is one particular way to implement the intervention $do({\mathcall X} = {\mathcall x})$ in the population.  In the second counterfactual population, a proportion $\mathbb{P}({\underline X_{t_0} ^{t_3}} = {\underline x_{t_0} ^{t_3}}^{*} \mid \mathcall{X} = \mathcall{x}^{*})$ of the individuals undergoes the intervention $do({\underline X_{t_0} ^{t_3}} = {\underline x_{t_0} ^{t_3}}^{*} )$ for any profile ${\underline x_{t_0} ^{t_3}}^{*}$ leading to ${\mathcall x}^*$,  which is one particular way to implement $do({\mathcall X} = {\mathcall x}^*)$ in the population. A similar, though stratum-specific, interpretation holds for the weighted average of Equation \eqref{Eq:SV_Sufficient_cond}. To recap, when $(Irrel_{\llbracket t_0, t_3 \rrbracket})$ does not hold, the weighted averages of Equations \eqref{Eq:SV_Sufficient_cond} and \eqref{Eq:SV_Sufficient_uncond} compare the expectations of the outcomes in the counterfactual worlds following two particular, and admittedly natural, implementations of the interventions $do({\mathcall X} = {\mathcall x})$ and $do({\mathcall X} = {\mathcall x}^*)$. As such, they correspond to causal effects of natural interest. 

However, focusing on the weighted averages of Equation \eqref{Eq:SV_Sufficient_uncond} for simplicity, it is important to keep in mind that the ``individual'' causal effects $ATE_{L}({\underline x_{t_0} ^{t_3}} ; {\underline x_{t_0} ^{t_3}}^{*})$, for two given profiles $\underline x_{t_0}^{t_3}$ and  $\underline x_{t_0}^{t_3*}$, involved in these weighted averages may be very different from one another when condition $(Irrel_{\llbracket t_0, t_3 \rrbracket})$ does not hold. In particular, when ``individual'' causal effects $ATE_{L}({\underline x_{t_0} ^{t_3}} ; {\underline x_{t_0} ^{t_3}}^{*})$ associated with large weights $\mathbb{P}({\underline X_{t_0} ^{t_3}} = {\underline x_{t_0} ^{t_3}} \mid \mathcall{X} = \mathcall{x}) \times \mathbb{P}( {\underline X_{t_0} ^{t_3}} = {\underline x_{t_0} ^{t_3}}^* \mid \mathcall{X} = \mathcall{x}^*)$, are very different from one another, the interpretation of the weighted average may be less straightforward. For illustration, consider the causal model ($L.0$) of Figure \ref{Fig:Example_Instantaneous} and its over-simplified counterpart ($S.0$), where $\mathcall{X} = X_{t_2}$ for some $t_2\in \llbracket 2; T\rrbracket $. In this case, we have $(Y^{\bar X_{t_2} = \bar x_{t_2}} \indep \bar X_{t_2})_{{L.0}}$ and  $(Y^{X_{t_2}=x_{t_2}}\indep X_{t_2})_{{S.0}}$, so that Theorem \ref{Theo_SV_Weak} ensures that 
\begin{eqnarray*}
\widetilde{ATE}_{S.0}(1; 0) & = &  \mathbb{E}\left( Y \mid X_{t_2} = 1 \right) - \mathbb{E}\left( Y \mid X_{t_2} = 0 \right),  \nonumber \\
& = & \sum_{\substack{\bar x_{t_2} \in \lbrace 0,1 \rbrace^{t_2} \\  \bar x_{t_2*} \in \lbrace 0,1 \rbrace^{t_2}}} \lbrace ATE_{L}\left( \bar x_{t_2} ; \bar x_{t_2*}\right) \times \mathbb{P}(\bar X_{t_2}  = \bar x_{t_2}  \mid X_{t_2} = 1)  \nonumber   \\[-0.9cm]
&& \hspace{4.5cm} \times \mathbb{P}(\bar X_{t_2} = \bar x_{t_2*}  \mid X_{t_2} = 0) \rbrace.
\end{eqnarray*}
In other words, $\widetilde{ATE}_{S.0}(1; 0)$ is the weighted sum of the longitudinal total effects that compare any possible pairs of exposure profiles up to time $t_2$, one of which terminating with $X_{t_2}=0$ and the other one terminating with $X_{t_2}=1$. In particular, terms like $ATE_{{L.0}} \left( (\textbf{0}_{t_2-1}, 1) ; (\textbf{1}_{t_2-1}, 0)\right)$, where $(\textbf{0}_{t_2-1}, 1)$ is an ``almost never exposed profile'' and $(\textbf{1}_{t_2-1}, 0)$ an ``almost always exposed profile'', have non-negative weights in  $\widetilde{ATE}_{{S.0}}(1;0)$, which complicates the interpretation of $\widetilde{ATE}_{{S.0}}(1;0)$.
On the other hand, the interpretation of $\widetilde{ATE}_{{S.0}}(1;0)$ is more straightforward if, for example, profiles $\bar x_{t_2-1}$ associated with large weights $\mathbb{P}( \bar X_{t_2 - 1} = \bar x_{t_2 - 1} \mid X_{t_2} = 1)$ correspond to globally more exposed profiles than profiles $\bar x^{*}_{t_2-1}$ associated with large weights $\mathbb{P}( \bar X_{t_2 - 1} = \bar x^{*}_{t_2 - 1} \mid X_{t_2} = 0)$. In particular, this is the case when the exposure is ``stable'', more precisely when $X_t = 1 \Rightarrow X_{t'}=1$ for all $t' \geq t$. Although this stability assumption is arguably rarely met in practice, it can be seen as a reasonable assumption (or approximation) for exposures such as obesity for instance. When it is satisfied, the only exposure profile that terminates with $x_{t_2} = 0$ is the 
``never-exposed profile'', $\bar x_{t_2} = \textbf{0}_{t_2}$, and, under model (\textit{L.0}), $\widetilde{ATE}_{{S.0}}(1;0)$ then reduces to
\begin{eqnarray*}
\sum _{i=0} ^{t_2-1} ATE_{L.0} \left( ( {\bf 0}_i , {\bf 1}_{t_2 -i}) ; {\bf 0}_{t_2} \right)   \times \mathbb{P}\left( \bar X_{t_2-1}=( {\bf 0}_i , {\bf 1}_{t_2 -i-1}) \mid X_{t_2}=1\right) . %\label{Eq:ATE_stable}
\end{eqnarray*}
If the true causal model is ($L.0$) and the over-simplified model is ($S.0$), the stability assumption guarantees that $\widetilde{ATE}_{{S.0}}(1;0)$ is a weighted sum of all the longitudinal causal effects comparing the ever-exposed profiles to the single never-exposed profile. Weights in the equation above are sensible as they correspond to the actual proportions of subjects with exposure profiles $( {\bf 0}_i , {\bf 1}_{t_2 -i})_{i \in \llbracket 0, t_2 - 1 \rrbracket}$ among the subpopulation of exposed individuals at time $t_2$. Therefore, $\widetilde{ATE}_{{S.0}}(1;0)$ can be regarded as a meaningful quantity under model (\textit{L.0}) of Figure \ref{Fig:Example_Instantaneous} if the stability assumption further holds. The fact that $\widetilde{ATE}_{{S.0}}(1;0)$ is a meaningful quantity under the stability assumption extends to the situation where a time-invariant observed confounder $W$ is added to model (\textit{L.0}). However, we shall stress that, if the confounder is time-varying, the conditions of Theorem \ref{Theo_SV_Weak} are generally not satisfied, and $\widetilde{ATE}_{{S.0}}(1;0)$ has usually no clear interpretation, even when both the exposure and confounder processes are stable.

\section{Numerical illustration }\label{sec:Illustr_SV}

In this Section, we empirically evaluate the magnitude of the bias between the quantity $\widetilde{ATE}_{S}(\mathcall{x} ; \mathcall{x}^*)$ estimated in practice when working under over-simplified causal models and the weighted averages of Equations \eqref{Eq:SV_Sufficient_cond} and \eqref{Eq:SV_Sufficient_uncond} when conditions $(T.Cond)$ and $(T.Uncond)$ of Theorem \ref{Theo_SV_Weak} are not satisfied. We consider a special case of model ($L.1$) where the time-varying confounder possibly affected by the exposure of interest only affects $Y$ through  ${\mathcall X}$ and ${\mathcall W}$, while the exposure of interest may have an effect on $Y$ beyond those through ${\mathcall X}$ and ${\mathcall W}$; see model ($L.6$) in Figure \ref{Fig:ATE_SV_example6_configuration}. Specifically, we set $T=5$, and consider binary variables $X_t$ and $W_t$ for $t=1, \ldots, 5$, and a continuous outcome $Y$, which are defined using the following system of structural equations. Denoting by $\xi_U$ and $f_U$ the exogenous variable and the structural function, respectively, attached to any particular variable $U$ of the model, we set $\xi_Y\sim {\mathcall N}(0, 1)$ while all other exogenous variables are univariate random variables uniformly distributed on  $[0, 1]$, and
\begin{eqnarray}
f_{W_1} \left( \xi_{W_1} \right) \hspace{-0.2cm} &=& \hspace{-0.2cm} \1 \left\lbrace \xi_{W_1} \leq c_{W_1} \right\rbrace, \label{Eq:Structural_Functions_Confounder_Affected} \\
f_{X_1} \left(  W_1, \xi_{X_1} \right) \hspace{-0.2cm} &=& \hspace{-0.2cm} \1 \left\lbrace \xi_{X_1} \leq {\rm expit} \left(  \alpha W_1 + c_{X_1} \right)  \right\rbrace \nonumber, \\
f_{W_t} \left( \bar W_{t-1}, \bar X_{t-1}, \xi_{W_t} \right)  \hspace{-0.2cm} &=& \hspace{-0.2cm} \1 \Big\lbrace \xi_{W_t} \leq {\rm expit}\big( \gamma \sum_{t' < t} W_{t'} + \rho \alpha X_{t-1} + c_{W_t} \big)\Big\rbrace, \forall t \in \llbracket 2 ; t_0 \rrbracket \nonumber,\\
f_{X_t} \left( \bar W_{t}, \bar X_{t-1}, \xi_{X_t} \right)  \hspace{-0.2cm} &=& \hspace{-0.2cm} \1 \Big\lbrace \xi_{X_t} \leq {\rm expit} \big(  \alpha \sum_{t' \leq t} W_{t'} + \beta \sum_{t' < t} X_{t'} + c_{X_t} \big) \Big\rbrace,   \forall t \in \llbracket 2 ; t_0 \rrbracket \nonumber, \\
f_Y(\mathcall{X}, \mathcall{W}, \xi_Y)  \hspace{-0.2cm} &=& \hspace{-0.2cm} \mu_0 + \mu_X \mathcall{X} - \mu_W \mathcall{W} + \frac{\mu_X}{T}\sum_{t = 1} ^T \theta^{T + 1 - t} X_t + \xi_Y \nonumber. 
\end{eqnarray}
Here ${\rm expit}(\cdot)$ denotes the sigmoid function, $\1\{\cdot\}$ denotes the indicator function, $\mathcall{X} = \mathbb{1}\left( \sum_{t=1} ^{T} X_t \geq 3\right)$ and $\mathcall{W} = \mathbb{1}\left( \sum_{t=1} ^{T} W_t \geq 3\right)$. Parameters $c_{W_1}$, $c_{X_1}$, $c_{W_t}$, and $c_{X_t}$ are set to values ensuring that the prevalence of $X_t$ and $W_t$ is about $0.2$ for all $t$. Parameter $\mu_W$ governs the strength of the effect of $(W_t)_{t\in \llbracket 1; T \rrbracket}$ on $Y$ through $\mathcall{W}$. In the same way, $\mu_X$ governs the strength of the effect of $(X_t)_{t\in \llbracket 1; T \rrbracket}$ on $Y$ through $\mathcall{X}$, but $X_t$ also has a ``direct'' effect on $Y$ when $\theta$ is non-zero. Parameter $\alpha$ governs the strength of the effect of $W_t$ on $X_{t'}$ for $t'\geq t$, while the strength of the effect of $X_t$ on $W_{t+1}$ is governed by the product $\rho\alpha$. Finally, parameter $\gamma$ governs the strength of the effect of $W_t$ on $W_{t'}$ for $t'\geq t$, and parameter $\beta$ governs the strength of the effect of $X_t$ on $X_{t'}$ for $t'\geq t$." %The case $\rho=0$ corresponds to the pure confounder case, where the confounder is not affected by the exposure, while  we get closer to the pure mediation setting as $\rho$ increases.

\begin{center}
\begin{figure}
\begin{center}
\begin{tikzpicture}[scale=0.78, auto,swap]
\node[var] (W2)at(2,0){$W_1$};
\node[var] (W3)at(4,0){$W_2$};
\node[var] (Wdot)at(5.5,0){$\dots$};
\node[var] (Wt)at(7,0){$W_{T}$};
\node[var] (W)at(8.5,0){$\mathcall{W}$};
\node[var] (X2)at(2,-1.5){$X_1$};
\node[var] (X3)at(4,-1.5){$X_2$};
\node[var] (Xdot)at(5.5,-1.5){$\dots$};
\node[var] (Xt)at(7,-1.5){$X_{T}$};
\node[var] (X)at(8.5,-1.5){$\mathcall{X}$};
\node[var] (Y)at(9.75,-0.75){$Y$};

\draw[edge] (X2)--(X3);
\draw[edge] (X2)--(W3);
\draw[edge] (X2).. controls (3.5, -0.85)..(Wt);
\draw[edge] (X3)--(Wt);

\draw[edge] (W2)--(W3);
\draw[edge] (W2)--(X2);
\draw[edge] (W2)--(X3);
\draw[edge] (W2).. controls (3.5, -0.75)..(Xt);
\draw[edge] (W3)--(Xt);

\draw[edge] (W3)--(X3);
\draw[edge] (W3)--(Wdot);
\draw[edge] (Wdot)--(Wt);
\draw[edge] (Wt)--(W);
\draw[edge] (Wt)--(Xt);

\draw[edge] (X3)--(Xdot);
\draw[edge] (Xdot)--(Xt);
\draw[edge] (Xt)--(X);

\draw[edge] (X)--(Y);
\draw[edge] (W)--(Y);

\draw[edge] (W2).. controls (5.1,0.75) ..(Wt);
\draw[edge] (W2).. controls (5,1.15) ..(W);

\draw[edge] (W3).. controls (5.1,0.5) ..(Wt);
\draw[edge] (W3).. controls (6.2,1.15) ..(W);

\draw[edge] (X2).. controls (5.1,-2.2) ..(Xt);
\draw[edge] (X2).. controls (5,-2.5) ..(X);
\draw[edge] (X2).. controls (5,-3.5) and (8.5,-3.5)..(Y);

\draw[edge] (X3).. controls (5.1,-1.9) ..(Xt);
\draw[edge] (X3).. controls (6.2,-2.5) ..(X);

\draw[edge] (X3).. controls (6.2,-3.1) and (7.9,-3.1)..(Y);

\draw[edge] (Xt).. controls (7.1,-2.5) and (7.9,-2.5)  ..(Y);

%\draw[edge] (W2).. controls (5,2.15) and (8.5,2.15)..(Y);
%\draw[edge] (W3).. controls (6.2,1.75) and (7.9,1.75)..(Y);
%\draw[edge] (Wt).. controls (7.1,1.15) and (7.9,1.15)  ..(Y);

\end{tikzpicture}
\vspace{0.1cm}

(\textit{L.6})
\end{center}

    \caption{(\textit{L.6}) Example of longitudinal model with  a time-varying confounder $(W_t)_{t \geq 1}$ possibly affected by the time-varying exposure of interest $(X_t)_{t \geq 1}$. These two processes potentially affect the outcome $Y$ through some summaries $\mathcall X$ and $\mathcall{W}$, respectively, which are assumed to be deterministic functions of $(X_t)_{t\geq 1}$ and $(W_t)_{t\geq 1}$, respectively.}
    \label{Fig:ATE_SV_example6_configuration}
\end{figure}
\end{center}

\begin{center}
\begin{figure}
\begin{center}
\includegraphics[scale=0.47]{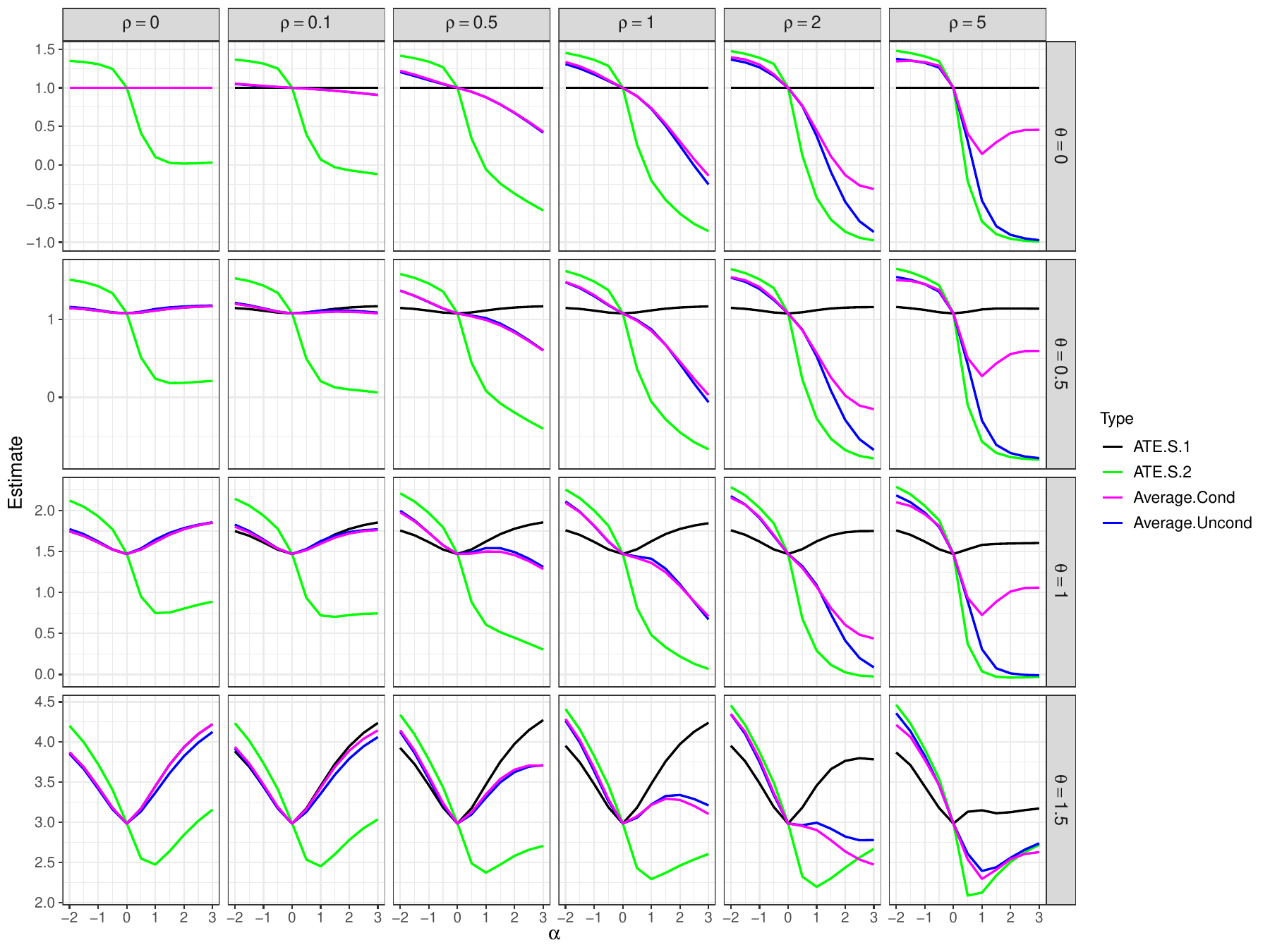} 
\end{center}
\caption{Values of $ATE_{S.1}\left(1;0 \right)$ (in black), $ATE_{S.2}\left(1;0\right)$ (in green), the weighted average of Equation \eqref{Eq:SV_Sufficient_cond} (in magenta) and the weighted average of Equation \eqref{Eq:SV_Sufficient_uncond} (in blue), under the causal model described in Equation \eqref{Eq:Structural_Functions_Confounder_Affected}.}
\label{Illu_Confounder_Affected}
\end{figure}
\end{center}

We compare the quantities estimated in practice when working under the over-simplified models ($S.1$) and ($S.2$), i.e.,  $\widetilde {ATE}_{S.1}\left(1;0\right)$  and $\widetilde{ATE}_{S.2}\left(1;0\right)$ respectively, with the weighted averages of Equations \eqref{Eq:SV_Sufficient_cond} and \eqref{Eq:SV_Sufficient_uncond} (with $\mathcall{x}=1$ and $\mathcall{x}^\star=0$). Monte-Carlo simulations based on samples of size $n=5\times10^6$ were used to approximate these 4 quantities. Figure \ref{Illu_Confounder_Affected} presents the results of the comparison for a set of combinations of parameters $\alpha\in[-2,3]$, $\beta= \vert \alpha \vert$, $\gamma=1$, $\theta\in \left\lbrace 0, 0.5, 1, 1.5 \right\rbrace$, $\rho\in \left\lbrace 0, 0.1, 0.5, 1, 2, 5 \right\rbrace$, $\mu_X = 1$ and  $\mu_W = 2$. The special case $\rho=0$ (first column in Figure \ref{Illu_Confounder_Affected}) corresponds to the scenario where the confounder is not affected by the exposure of interest (pure confounding), in which case condition ($T.Cond$) is satisfied if the over-simplified causal model is ($S.1$) in Figure \ref{fig:example_oversimplified_XMW}, so $\widetilde{ATE}_{S.1}$ equals the weighted average of Equation \eqref{Eq:SV_Sufficient_cond}. If, in addition, $\theta=0$ (in which case $\bar X_T$ affects $Y$ through $\mathcall{X}$ only), then versions of $\mathcall{X}$ are irrelevant, and $\widetilde{ATE}_{S.1}(1, 0) = ATE_{L.6}( \bar x_{t_0}^{t_3} ; {\bar x_{t_0}^{t_3*}})$  for any $\bar x_{t_0}^{t_3}$ and ${\bar x_{t_0}^{t_3*}}$ leading to $\mathcall{X} = 1$ and $\mathcall{X} = 0$, respectively. Therefore, $\widetilde{ATE}_{S.1}(1, 0)$ is also equal to the weighted average of Equation \eqref{Eq:SV_Sufficient_uncond}.  This is also true  when $\alpha = 0$, irrespective of the value of $\theta$. Indeed, this case corresponds to the scenario with no mediation and no confounding. Then, $\widetilde {ATE}_{S.1}\left(1;0\right) = \widetilde {ATE}_{S.2}\left(1;0\right)$, and these two quantities equal the two weighted averages of Equations \eqref{Eq:SV_Sufficient_cond} and  \eqref{Eq:SV_Sufficient_uncond}. In particular, versions are again irrelevant if, in addition, $\theta=0$, in which case we have  $\widetilde {ATE}_{S.1}\left(1;0\right) = \widetilde {ATE}_{S.2}\left(1;0\right) =  ATE_{L.6}\left( \bar x_{t_0}^{t_3} ; {\bar x_{t_0}^{t_3*}}\right)$ for any $\bar x_{t_0}^{t_3}$ and ${\bar x_{t_0}^{t_3*}}$ leading to $\mathcall{X} = 1$ and $\mathcall{X} = 0$, respectively. For all other combinations of parameters, both $\widetilde {ATE}_{S.1}\left(1;0\right)$ and $\widetilde {ATE}_{S.2}\left(1;0\right)$ differ from  the weighted average of Equations \eqref{Eq:SV_Sufficient_cond} and \eqref{Eq:SV_Sufficient_uncond}. When $\rho = 0.1$, $(W_t)_{t\geq 1}$ mostly acts as a confounder (and not so much as a mediator), and the difference between $\widetilde {ATE}_{S.1}\left(1;0\right)$ and the weighted averages is generally limited. But the difference between $\widetilde {ATE}_{S.1}\left(1;0\right)$ and the weighted averages of Equations \eqref{Eq:SV_Sufficient_cond} and  \eqref{Eq:SV_Sufficient_uncond} can be substantial for larger values of $\rho$. In particular, because the effect of ${\mathcall W}$ on $Y$ is $-\mu_W$, the indirect effect of the exposure process through $(W_t)_t$ is negative for positive $\alpha$, so that the weighted averages can be negative for some combinations of values for $\rho$, $\alpha$ and $\theta$, while $\widetilde {ATE}_{S.1}\left(1;0\right)$ is consistently positive. For large values of $\rho$,  $(W_t)_{t\geq 1}$ mostly acts as a mediator, and the difference between $\widetilde {ATE}_{S.2}\left(1;0\right)$ and the weighted average of Equation \eqref{Eq:SV_Sufficient_uncond} is generally limited. 

\begin{center}
\begin{figure}
\begin{center}
\includegraphics[scale=0.48]{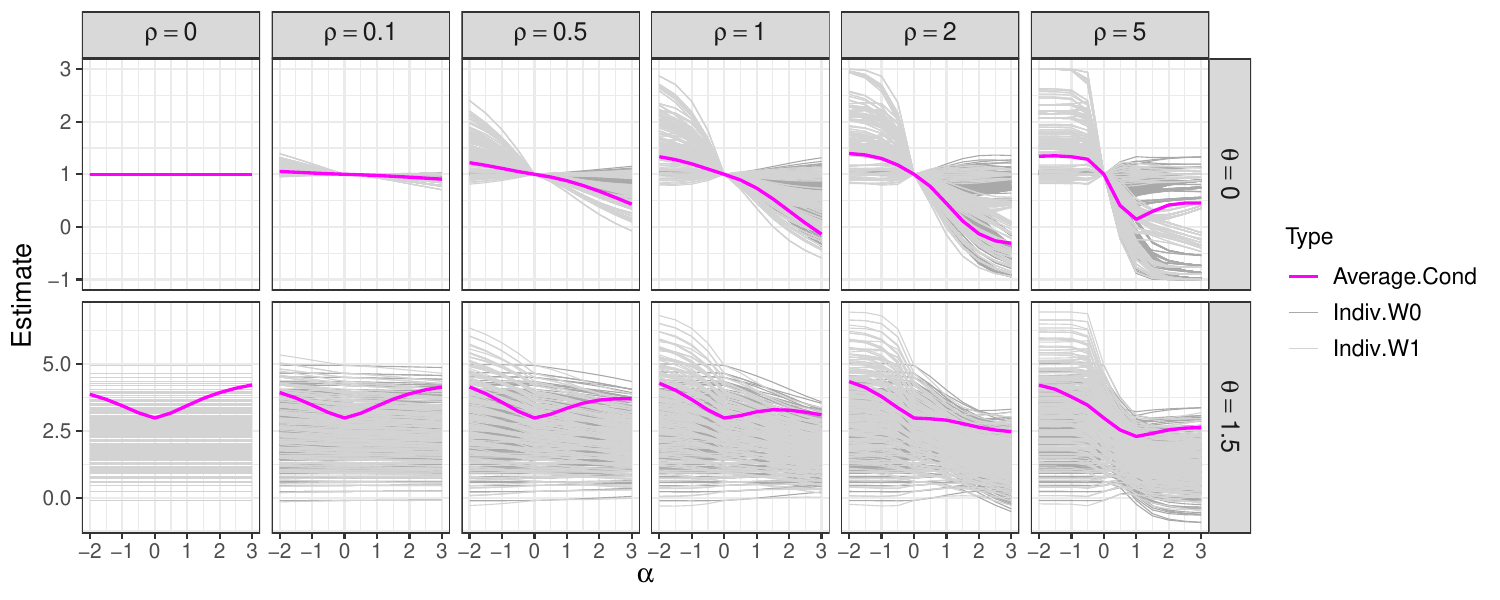} 
\end{center}
\caption{Values of $ATE_{L_{\mid \mathcall{W} = 0}}$ (in dark gray), of $ATE_{L_{\mid \mathcall{W} = 1}}$ (in gray), for each couple of exposure profiles leading to $\mathcall{X}= 1$ and $\mathcall{X} =  0$, and of the weighted average \eqref{Eq:SV_Sufficient_cond} (in magenta), under the causal model described in Equation \eqref{Eq:Structural_Functions_Confounder_Affected}.}
\label{Illu_CondAv}
\begin{center}
\includegraphics[scale=0.5]{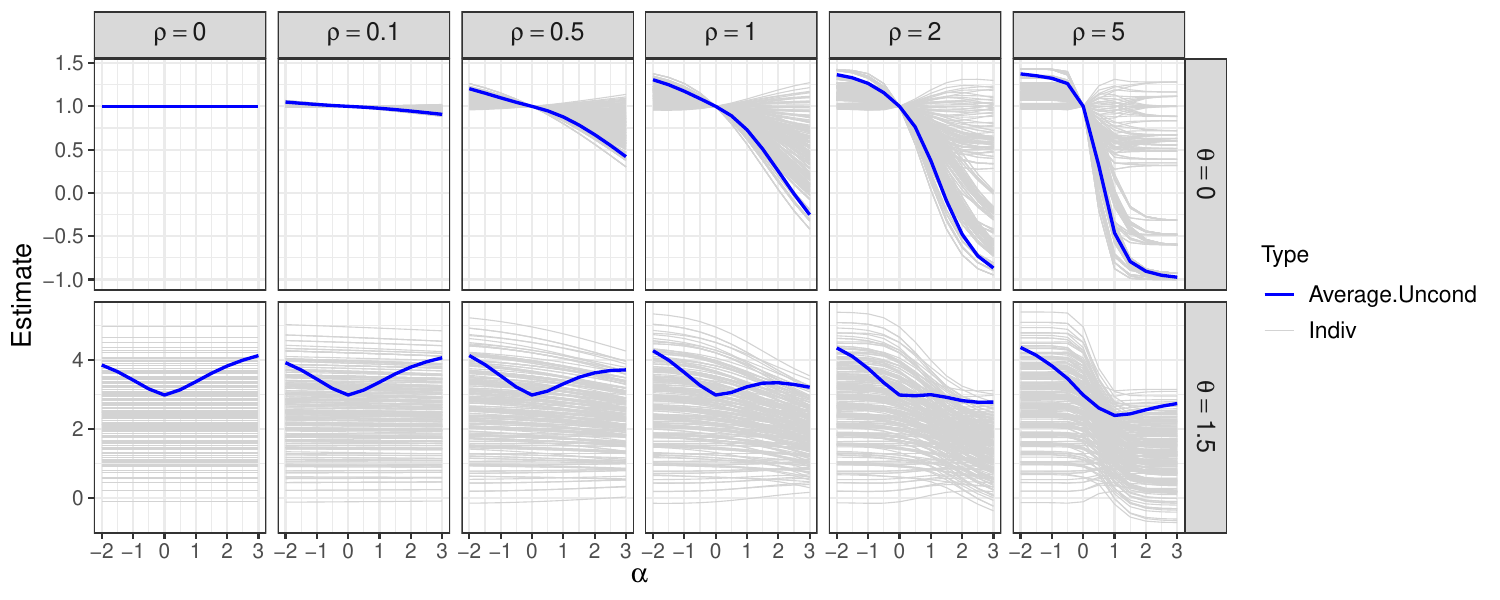} 
\end{center}
\caption{Values of $ATE_{L}(\bar x_{t_0} ; \bar x_{t_0}^{*})$ (in gray) for each couple of exposure profiles leading to $\mathcall{X}= 1$ and $\mathcall{X} =  0$, and of the weighted average \eqref{Eq:SV_Sufficient_uncond} (in blue), under the causal model described in Equation \eqref{Eq:Structural_Functions_Confounder_Affected}.}
\label{Illu_UncondAv}
\end{figure}
\end{center}

Finally, let us turn our attention to the individual causal effects $ATE_{L_{\mid \mathcall{W} = 1}}(\bar x_{t_0};$ $\bar x_{t_0}^{*})$ and $ATE_{L_{\mid \mathcall{W} = 0}}(\bar x_{t_0};$ $\bar x_{t_0}^{*})$ and $ATE_{L}\left( \bar x_{t_0} ; \bar x_{t_0}^{*}\right)$ involved in the weighted averages of Equations \eqref{Eq:SV_Sufficient_cond} and \eqref{Eq:SV_Sufficient_uncond}. Figures \ref{Illu_CondAv} and \ref{Illu_UncondAv} display these individual causal effects for some particular combinations of the parameters of our model. As shown in Figure \ref{Illu_UncondAv}, the values of the individual causal effects $ATE_{L}\left( \bar x_{t_0} ; \bar x_{t_0}^{*}\right)$ are relatively homogeneous for negative values of $\alpha$ and $\rho\leq 2$. In particular, versions of treatment $\mathcall{X}$ are irrelevant and the individual causal effects $ATE_{L}\left( \bar x_{t_0} ; \bar x_{t_0}^{*}\right)$ are all equal when $\theta = 0$, $\rho=0$ and/or $\alpha=0$, (in this case, they are also equal to the individual causal effects $ATE_{L_{\mid \mathcall{W} = 1}}(\bar x_{t_0};$ $\bar x_{t_0}^{*})$, $ATE_{L_{\mid \mathcall{W} = 0}}(\bar x_{t_0};$ $\bar x_{t_0}^{*})$). The values of the individual causal effects are more heterogeneous for other combinations of the parameters, especially when both $\rho$ and $\alpha$ are large. In these situations, the weighted averages of Equation \eqref{Eq:SV_Sufficient_uncond} (and therefore $\widetilde{ATE}_{S.2}(1;0)$ too) and Equation \eqref{Eq:SV_Sufficient_cond} may have to be interpreted with more caution. But again, inspecting the weights involved in these weighted averages is instructive. For example, consider the case where $\theta = 0$, $\alpha = 3$ and $\rho = 5$ in Figure \ref{Illu_UncondAv}. Although the individual terms are very heterogeneous, three of them contribute 96\% of the total weights:  $ATE_{L}\big( {\bf 1}_{5} ; {\bf 0}_{5} \big) \approx -1$, $ATE_{L}\big( ( {\bf 0}_1 , {\bf 1}_{4}) ; {\bf 0}_{5} \big)\approx -1$,  and $ATE_{L}\big( ( {\bf 1}_1 , {\bf 0}_1, {\bf 1}_{3}) ; {\bf 0}_{5} \big)\approx -1$, with respective weights 0.835, 0.106 and 0.025 (the forth largest weight is only 0.001). In other words, although many different profiles $\bar x_{t_0}$ and $\bar x_{t_0}^{*}$ can lead to $\mathcall{X}=1$ and $\mathcall{X}=0$, respectively, and the corresponding causal effects $ATE_{L}(\bar x_{t_0};$ $\bar x_{t_0}^{*})$ can be very heterogeneous, mostly one profile (${\bf 0}_{5}$) is observed to lead to $\mathcall{X}=0$ and mostly three profiles (${\bf 1}_{5}$, $( {\bf 0}_1 , {\bf 1}_{4})$ and $( {\bf 1}_1 , {\bf 0}_1, {\bf 1}_{3})$) are observed to lead to $\mathcall{X}=1$ in this particular example. Moreover, the three corresponding causal effects $ATE_{L}\big( {\bf 1}_{5} ; {\bf 0}_{5} \big) $, $ATE_{L}\big( ( {\bf 0}_1 , {\bf 1}_{4}) ; {\bf 0}_{5} \big)$,  and $ATE_{L}\big( ( {\bf 1}_1 , {\bf 0}_1, {\bf 1}_{3}) ; {\bf 0}_{5} \big)$ happen to all have very similar values. This can therefore be seen as an example where the weighted average has a relatively clear meaning, and so has $\widetilde{ATE}_{S.2}\left(1;0\right)$, since very little bias was observed when working under the over-simplified model (S.2) in this case; see Figure  \ref{Illu_Confounder_Affected}. 

\section{Discussion}\label{sec:Discu}

The longitudinal nature of risk factors is often overlooked in epidemiology. In this article, we investigated whether causal effects derived under over-simplified models that ignore the time-varying nature of exposures could still be related to causal effects of potential interest.  We focused on the general setting, where the available data concerns summaries $(\mathcall X, \mathcall Z)$ of the full exposure profiles $(\bar X_T, \bar Z_T)$, instead of the full exposure profile themselves, where $\mathcall X$ and $\mathcall Z$ correspond to deterministic functions of $(\bar X_T, \bar Z_T)$. This general framework includes the special case where $\mathcall X = X_{t_2}$ and $\mathcall Z = Z_{t_2}$, for some time $t_2 \in \llbracket 1;T \rrbracket$ ({\it e.g.}, inclusion time in the study). Under the conditions of Theorem \ref{Theo_SV_Weak}, the quantity estimated in practice expresses as a weighted average of longitudinal causal effects. But, first, these conditions are very restrictive, and second, even when they are met, the interpretation of the weighted averages is not always straightforward. Therefore, and unsurprisingly, our results are mostly negative, as they state that the quantity estimated in practice when working under over-simplified causal models has generally no clear interpretation in terms of longitudinal causal effects of interest, except under very simple longitudinal causal models.

The bias resulting from this kind of simplification of the true causal model has been less studied and acknowledged than the more standard confounding, collider and selection biases \citep{Greenland_Bias2003, DingMiratrix_Bias2015, HernanRobins_SelectionBias2004, Hernan_HR2010}. Overall, our results are consistent with, and complete, a few previous results, which already stressed the need for appropriate statistical methods applied to repeated measurements of exposures when the true causal model is longitudinal, and suggested that overlooking the time-varying nature of exposures would generally lead to unreliable causal effect estimates \citep{G_Formula, article2, article1}. Other related works include those on data coarsening, as that of \cite{TargetedEHRData_Sofrygin2019}, who empirically showed that while reducing the computational cost of the analyses, partitioning the follow-up into coarse intervals may lead to invalid inference. The study of the impact of the discretization of a continuous-time causal model constitutes an interesting lead for future research.

Just as in the presence of unobserved confounders, we encourage analysts to consider the full DAG representing the  causal model rather than the over-simplified one, even when only summary measures, or measures at single time point, are available. This could help them to identify possible biases and not to over-interpret the estimated quantity. This could also be used to develop sensitivity analyses to appreciate the magnitude of the corresponding bias. Moreover, in rare cases, general results on the identifiability of causal effects in the presence of unobserved variables could be applied \citep{TianPearl2002, ShpitserPearl2006a, HuangValtorta2006} to determine whether particular longitudinal causal effects of interest can be identified from the available data. For example, consider the causal model $(L.fd)$ of Figure \ref{fig:FrontDoor}: in this particular case,  $ATE_L^{(T)}(1;0) = \mathbb{E}_{L}\left( Y^{X_{T}=1} - Y^{X_{T}=0}\right)$ can be identified using the front-door criterion \citep{pearl_book}. In other words, $ATE_L^{(T)}(1;0)$ can be estimated with data on $X_{T}$, $M_{T}$ and $Y$ only. Conversely, if the analyst focuses on the observed variables only, and works under the simplified model $(S.fd)$, inference would be based on $ATE_S^{(T)}(1;0) = \mathbb{E}_{S}\left( Y^{X_{T}=1} - Y^{X_{T}=0}\right)$, which would be identified as $\widetilde{ATE}_S^{(T)}(1;0) := \mathbb{E}\left( Y \mid X_{T}=1\right)  -  \mathbb{E}\left( Y \mid  X_{T}=0\right)$. Moreovoer, Theorem \ref{Theo_SV_Weak} applies in this case and ensures that
\begin{eqnarray*}
\widetilde{ATE}_S^{(T)}(1;0) &=& \sum_{\substack{\bar x_{T-1} \in \lbrace 0,1 \rbrace^{T-1} \\  \bar x_{T-1*} \in \lbrace 0,1 \rbrace^{T-1}}} \lbrace ATE_{L}\left( (\bar x_{T-1},1) ; (\bar x_{T-1}^*,0)\right)\\[-0.9cm]
&& \hspace{4.5cm}  \times \mathbb{P}(\bar X_{T-1}  = \bar x_{T-1}  \mid X_{T} = 1)  \nonumber   \\
&& \hspace{4.5cm} \times \mathbb{P}(\bar X_{T-1} = \bar x_{T-1}^{*}  \mid X_{T} = 0) \rbrace.
\end{eqnarray*}
However, this weighted average differs from $ATE_L^{(T)}(1;0)$, and is generally more complicated to interpret (unless some stability assumption holds for example).

\begin{center}
\begin{figure}
\begin{minipage}[c]{0.72\linewidth}
\begin{tikzpicture}[scale=0.87, auto,swap]
\node[var] (X1)at(0,0){$X_1$};
\node[var] (M1)at(1.25,0){$M_1$};
\node[var] (dots)at(2.35,0){$\dots$};
\node[var] (Xt)at(3.75,0){$X_{T-1}$};
\node[var] (Mt)at(5.5,0){$M_{T-1}$};
\node[var] (Xtt)at(6.95,0){$X_{T}$};
\node[var] (Mtt)at(8.15,0){$M_{T}$};
\node[var] (Y)at(9.25,0){$Y$};
\draw[edge] (X1)--(M1);
\draw[edge] (M1)--(dots);
\draw[edge] (dots)--(Xt);
\draw[edge] (Xt)--(Mt);
\draw[edge] (Mt)--(Xtt);
\draw[edge] (Xtt)--(Mtt);
\draw[edge] (Mtt)--(Y);
\draw[edge] (Mt).. controls (7.25,0.5) ..(Y);
\draw[edge] (M1).. controls (6.5,1) ..(Y);
\end{tikzpicture}
\end{minipage}\hfill
\begin{minipage}[c]{0.27\linewidth}
\begin{center}
\begin{tikzpicture}[scale=0.95, auto,swap]
\node[var] (Xtt)at(6.95,0){$X_{T}$};
\node[var] (Mtt)at(8.15,0){$M_{T}$};
\node[var] (Y)at(9.25,0){$Y$};
\draw[edge] (Xtt)--(Mtt);
\draw[edge] (Mtt)--(Y);
\end{tikzpicture}
\end{center}
\end{minipage}

\vspace{0.2cm}

\begin{minipage}[c]{0.75\linewidth}
\begin{center}
(\textit{L.fd})
\end{center}
\end{minipage}\hfill
\begin{minipage}[c]{0.23\linewidth}
\begin{center}
(\textit{S.fd})
\end{center}
\end{minipage}
\caption{An example where the front door criterion applies and allows the identification of $ATE_L^{(T)}(1;0) = \mathbb{E}_{L}\left( Y^{X_{T}=1} - Y^{X_{T}=0}\right)$. Conversely, the quantity estimated when ignoring the time-varying nature of the exposures, and working under the over-simplified causal model ($S.fd$), usually does not have a clear causal meaning.}
\label{fig:FrontDoor}
\end{figure}
\end{center}

A large amount of prospectively collected repeated measures is being made available for a number of exposures through electronic health records and their linkage to biobanks \citep{BiobankRef_Beesley2020}. However, their analysis raises other challenges, including those pertaining to selection bias \citep{Agnielk1479, EHR_Beesley2020_medRxiv, RefConclu, EHR_Beesley2020}. Prospectively collecting repeated measures of exposures, including biomarkers and -omics data, in well designed cohort studies, as a few studies already did \citep{koges2017}, would be valuable to  assess the causal effects of exposures on health-related outcomes.

\section*{Acknowledgments}

The authors are grateful to Stijn Vansteelandt for insightful comments on preliminary versions of this article.

\section*{Disclaimers}

Where authors are identified as personnel of the International Agency for Research on Cancer / World Health Organization, the authors alone are responsible for the views expressed in this article and they do not necessarily represent the decisions, policy or views of the International Agency for Research on Cancer / World Health Organization.

\vspace{0.8cm}

\appendix {\noindent\Large{\bf Appendices}}

\section{Proof of Theorem \ref{Theo_SV_Weak}}\label{Proof:Theo_SV_Weak}

Consider a longitudinal model ($L$) and assume that the only available data regarding the exposure of interest consists in $\mathcall X$, which is a deterministic function of $\underline X_{t_1} ^{t_2}$, with $1 \leq t_1 \leq t_2 \leq T$. Let $\mathcall{x}$ and $\mathcall{x}^{*}$, $\mathcall{x} \neq \mathcall{x}^{*}$, be two given possible values of $\mathcall X$, and assume that there exists some $\mathcall{W} \subset \mathcall{Z}$, taking its values in some space $\Omega_{\mathcall{W}}$, such that  $ 0<\mathbb{P}(\mathcall{X} = \mathcall{x} \mid \mathcall{W} = \mathcall{w}) < 1$ and $0 < \mathbb{P}(\mathcall{X} = \mathcall{x}^{*} \mid \mathcall{W} = \mathcall{w})< 1$, for all $\mathcall{w}$ such that $\mathbb{P}(\mathcall{W} = \mathcall{w}) >0$. Now, consider an over-simplified model ($S$) and assume that $(Y^{\mathcall{X} = \mathcall{x}} \indep \mathcall{X} \mid \mathcall{W})_S$. Then, following usual arguments of causal inference \citep{pearl2009statsurvey, robins1986, rosenbaum1983}, the causal effect of interest, $ATE_{S} \left(\mathcall{x} ; \mathcall{x}^{*}\right)   :=\mathbb{E}_{S}\left( Y^{\mathcall{X} = \mathcall{x}} - Y^{\mathcall{X} = \mathcall{x}^{*}}\right)$, would be estimated under this over-simplified model $(S)$ as
\begin{eqnarray*}
\widetilde{ATE}_{S} \left(\mathcall{x} ; \mathcall{x}^{*}\right) &=& \sum_{\mathcall{w} \in \Omega_{\mathcall{W}}} \left[ \mathbb{E}\left( Y \mid \mathcall{W} = \mathcall{w}, \mathcall{X} = \mathcall{x} \right) - \mathbb{E}\left( Y \mid \mathcall{W} = \mathcall{w}, \mathcall{X} = \mathcall{x}^{*} \right) \right] \nonumber \\[-0.3cm]
&& \hspace{0.95cm} \times \mathbb{P}(\mathcall{W}=\mathcall{w}).
\end{eqnarray*}
Now assume that there exists some $t_0 \in \llbracket 1, t_1 \rrbracket$, some $t_3 \in \llbracket t_2, T \rrbracket$, such that $(Y^{\underline X_{t_0}^{t_3} = \underline x_{t_0} ^{t_3}} \indep \underline X_{t_0}^{t_3} \mid \mathcall{W})_L$. Let $\mathcall{w}$ be any given possible value of $\mathcall{W}$ such that $\mathbb{P}(\mathcall{W} = \mathcall{w}) >0$, and $\underline x_{t_0} ^{t_3}$ and $\underline x_{t_0} ^{t_3*}$ in $\left\lbrace 0, 1 \right\rbrace ^{t_3 - t_0 + 1}$ be any two possible profiles of $\underline X_{t_0}^{t_3}$ leading to $\mathcall{X} = \mathcall{x}$ and $\mathcall{X} = \mathcall{x}^*$, respectively, and  such that $\mathbb{P}(\underline X_{t_0} ^{t_3} = \underline x_{t_0} ^{t_3} \mid \mathcall{X} = \mathcall{x}, \mathcall{W} = \mathcall{w}) >0$ and $\mathbb{P}(\underline X_{t_0} ^{t_3} = \underline x_{t_0} ^{t_3*} \mid \mathcall{X} = \mathcall{x}^*, \mathcall{W} = \mathcall{w}) >0$. It follows that $ \mathbb{P}(\underline X_{t_0} ^{t_3} = \underline x_{t_0} ^{t_3} \mid \mathcall{W} = \mathcall{w}) > 0$ and $\mathbb{P}(\underline X_{t_0} ^{t_3} = \underline x_{t_0} ^{t_3*} \mid \mathcall{W} = \mathcall{w})>0$. Next, usual arguments of causal inference \citep{pearl2009statsurvey, robins1986, rosenbaum1983} yield
\begin{eqnarray*}
ATE_{L_{\mid \mathcall{W} = \mathcall{w}}} \left( \underline x_{t_0}^{t_3} ; \underline x^{t_3*}_{t_0}\right)
\hspace{-0.25cm} &=& \hspace{-0.55cm} \sum_{\mathcall{w} \in \Omega_{\mathcall{W}}} \mathbb{E}_{L}\left( Y^{\underline X_{t_0}^{t_3} = \underline x_{t_0}^{t_3} }   -  Y^{\underline X_{t_0}^{t_3} =  \underline x^{t_3 *}_{t_0} } \mid  \mathcall{W} = \mathcall{w} \right), \\
\hspace{3cm} &=& \hspace{-0.55cm} \sum_{\mathcall{w} \in \Omega_{\mathcall{W}}} \Big[ \mathbb{E}_{L} \Big( Y^{\underline X_{t_0} ^{t_3} = \underline x_{t_0}^{t_3} } \mid \mathcall{W} = \mathcall{w},  {\underline X_{t_0}^{t_3} = \underline x_{t_0}^{t_3} } \Big)  \\[-0.3cm]
&& \hspace{1.05cm}  -  \mathbb{E}_{L} \Big( Y^{\underline X_{t_0}^{t_3} =  \underline x^{t_3*}_{t_0} } \mid \mathcall{W} = \mathcall{w}, {\underline X_{t_0}^{t_3} =  \underline x^{t_3*}_{t_0} } \Big) \Big] ,\\
%\end{eqnarray*}
%\begin{eqnarray*}
\hspace{2.8cm}&=& \hspace{-0.55cm} \sum_{\mathcall{w} \in \Omega_{\mathcall{W}}} \Big[ \mathbb{E} \Big( Y \mid \mathcall{W} = \mathcall{w},  {\underline X_{t_0}^{t_3} = \underline x_{t_0}^{t_3} } \Big)   -  \mathbb{E} \Big( Y \mid \mathcall{W} = \mathcall{w}, {\underline X_{t_0}^{t_3} =  \underline x^{t_3*}_{t_0} } \Big) \Big].
\end{eqnarray*}

Because $\underline X_{t_0}^{t_3}$ \textit{d}-separates $\mathcall{X}$ and $\mathcall{W}$ under model ($L$)  \citep{pearl_book, Verma_Pearl88}, we have, for any any $\mathcall{w}$ in $\Omega_{\mathcall{W}}$ and any $\underline x_{t_0}^{t_3}$ in $\lbrace 0 , 1\rbrace ^{t_3 - t_0 +1}$ such that $\mathbb{P}(\underline X_{t_0} ^{t_3} = \underline x_{t_0} ^{t_3} \mid  \mathcall{W} = \mathcall{w}) >0$, 
\begin{eqnarray*}
\mathbb{E} \Big( Y \mid \mathcall{W} = \mathcall{w},  {\underline X_{t_0}^{t_3} = \underline x_{t_0}^{t_3} } \Big) &=& \mathbb{E} \Big( Y \mid \mathcall{W} = \mathcall{w}, \mathcall{X} = \mathcall{x}, {\underline X_{t_0}^{t_3} = \underline x_{t_0}^{t_3} } \Big),
\end{eqnarray*}
with $\mathcall{x}$ corresponding to the value taken by $\mathcall{X}$ when $\underline X_{t_0}^{t_3} = \underline x_{t_0}^{t_3}$. In other respect, we have
\begin{eqnarray*}
\mathbb{E}\left( Y \mid \mathcall{W} = \mathcall{w}, \mathcall{X} = \mathcall{x} \right) &=& \sum_{\underline x_{t_0}^{t_3} \in \left\lbrace 0, 1 \right\rbrace ^{t_3 - t_0 + 1}} \mathbb{E}\left( Y \mid \mathcall{W} = \mathcall{w}, \mathcall{X} = \mathcall{x}  , {\underline X_{t_0}^{t_3} = \underline x_{t_0}^{t_3} } \right) \\[-0.6cm]
 && \hspace{2.75cm} \times \mathbb{P}(\underline X_{t_0}^{t_3} = \underline x_{t_0}^{t_3} \mid \mathcall{W} = \mathcall{w}, \mathcall{X} = \mathcall{x} ) , \\
 &=& \sum_{\underline x_{t_0}^{t_3} \in \left\lbrace 0, 1 \right\rbrace ^{t_3 - t_0 + 1}} \mathbb{E}\left( Y \mid \mathcall{W} = \mathcall{w}, {\underline X_{t_0}^{t_3} = \underline x_{t_0}^{t_3} } \right) \\[-0.6cm]
 && \hspace{2.75cm} \times \mathbb{P}(\underline X_{t_0}^{t_3} = \underline x_{t_0}^{t_3} \mid \mathcall{W} = \mathcall{w}, \mathcall{X} = \mathcall{x} ) , \\
&=& \sum_{\underline x_{t_0}^{t_3}\in \left\lbrace 0, 1 \right\rbrace ^{t_3 - t_0 + 1}} \mathbb{E}_{L} \Big( Y^{\underline X_{t_0}^{t_3} = \underline x_{t_0}^{t_3} } \mid \mathcall{W} = \mathcall{w}   \Big) \\[-0.6cm]
 && \hspace{2.75cm} \times \mathbb{P}(\underline X_{t_0}^{t_3} = \underline x_{t_0}^{t_3} \mid \mathcall{W} = \mathcall{w}, \mathcall{X} = \mathcall{x} ),
\end{eqnarray*}
where the second equality comes from the fact that $\underline X_{t_0}^{t_3} = \underline x_{t_0}^{t_3} \Rightarrow \mathcall{X} = \mathcall{x}$, for any $\underline x_{t_0}^{t_3}$ such that $\mathbb{P}(\underline X_{t_0}^{t_3} = \underline x_{t_0}^{t_3} \mid \mathcall{W} = \mathcall{w}, \mathcall{X} = \mathcall{x})\neq 0$. This finally yields
\begin{eqnarray*}
\widetilde{ATE}_{S}(\mathcall{x}; \mathcall{x}^*) \hspace{-0.25cm} & = & \hspace{-0.55cm} \sum_{\mathcall{w} \in \Omega_{\mathcall{W}}} \hspace{-0.3cm} \sum_{\substack{\underline x_{t_0}^{t_3}, \\ \underline x^{t_3*}_{t_0} \in \left\lbrace 0, 1 \right\rbrace ^{t_3 - t_0 + 1}}} \hspace{-0.8cm} \lbrace ATE_{L_{\mid \mathcall{W} = \mathcall{w}}}\left( \underline x_{t_0}^{t_3} ; \underline x^{t_3*}_{t_0}\right) \times \mathbb{P}(\underline X_{t_0}^{t_3} = \underline x_{t_0}^{t_3} \mid \mathcall{X} = \mathcall{x}, \mathcall{W} = \mathcall{w})  \nonumber   \\[-1.1cm]
&& \hspace{5.4cm} \times \mathbb{P}(\underline X_{t_0}^{t_3} = \underline x_{t_0}^{t_3*} \mid \mathcall{X} = \mathcall{x}^{*}, \mathcall{W} = \mathcall{w})\\
&& \hspace{5.4cm} \times \mathbb{P}(\mathcall{W} = \mathcall{w})\rbrace,
\end{eqnarray*}
where the sums are over all $\bar x_{t_0}^{t_3}$ and $\bar x_{t_0}^{t_3*}$ in $\lbrace 0 , 1 \rbrace^{t_3 - t_0 + 1}$ such that $\mathbb{P}(\underline X_{t_0}^{t_3} = \underline x_{t_0}^{t_3} \mid \mathcall{W} = \mathcall{w}, \mathcall{X} = \mathcall{x})$ and $\mathbb{P}(\underline X_{t_0}^{t_3} = \underline x_{t_0}^{t_3*} \mid \mathcall{W} = \mathcall{w}, \mathcall{X} = \mathcall{x}^*)$, respectively, are not null.\\

The proof of the result under condition ($T.Uncond$) follows from similar, but simpler, arguments and is therefore omitted.

\end{document}